\documentclass[twocolumn]{aastex62}
\usepackage[utf8]{inputenc}
\usepackage{xspace}
\usepackage{amsmath}
\usepackage{url}
\usepackage{graphicx}

\shorttitle{GBNCC Pulsar Census} 
\shortauthors{McEwen et al.}

\begin{document}

\title{The Green Bank North Celestial Cap Pulsar Survey. V. Pulsar Census and Survey Sensitivity}

\author[0000-0001-5481-7559]{A.~E.~McEwen}
\affiliation{Center for Gravitation, Cosmology, and Astrophysics, 
Department of Physics, University of Wisconsin-Milwaukee, P.O. Box 413, Milwaukee, WI 53201, USA}

\author[0000-0002-6730-3298]{R.~Spiewak}
\affiliation{Centre for Astrophysics and Supercomputing, 
Swinburne University of Technology, P.O. Box 218, Hawthorn, VIC 3122, Australia}

\author[0000-0002-1075-3837]{J.~K.~Swiggum}
\affiliation{Center for Gravitation, Cosmology, and Astrophysics, 
Department of Physics, University of Wisconsin-Milwaukee, P.O. Box 413, Milwaukee, WI 53201, USA}

\author[0000-0001-6295-2881]{D.~L.~Kaplan}
\affiliation{Center for Gravitation, Cosmology, and Astrophysics, 
Department of Physics, University of Wisconsin-Milwaukee, P.O. Box 413, Milwaukee, WI 53201, USA}

\author[0000-0001-5645-5336]{W.~Fiore}
\affiliation{Department of Physics and Astronomy, West Virginia University, Morgantown, WV 26506, USA}
\affiliation{Center for Gravitational Waves and Cosmology,
West Virginia University, Chestnut Ridge Research
Building, Morgantown, WV 26505}

\author{G.~Y.~Agazie}
\affiliation{Department of Physics and Astronomy, West Virginia University, Morgantown, WV 26506, USA}
\affiliation{Center for Gravitational Waves and Cosmology, 
West Virginia University, Chestnut Ridge Research 
Building, Morgantown, WV 26505}

\author[0000-0003-4046-884X]{H.~Blumer}
\affiliation{Department of Physics and Astronomy, West Virginia University, Morgantown, WV 26506, USA}
\affiliation{Center for Gravitational Waves and Cosmology, 
West Virginia University, Chestnut Ridge Research 
Building, Morgantown, WV 26505}

\author[0000-0002-3426-7606]{P.~Chawla}
\affiliation{Department of Physics \& McGill Space Institute, McGill University, 3600 University Street, Montreal, QC H3A 2T8, Canada}

\author[0000-0002-2185-1790]{M.~DeCesar}
\affiliation{Department of Physics, 730 High St., Lafayette College, Easton, PA 18042, USA}

\author[0000-0001-9345-0307]{V.~M.~Kaspi}
\affiliation{Department of Physics \& McGill Space Institute, McGill University, 3600 University Street, Montreal, QC H3A 2T8, Canada}

\author[0000-0001-8864-7471]{V.~I.~Kondratiev}
\affiliation{ASTRON, the Netherlands Institute for Radio Astronomy, 
Oude Hoogeveensedijk 4, 7991 PD Dwingeloo, The Netherlands}
\affiliation{Astro Space Centre, Lebedev Physical Institute, 
Russian Academy of Sciences, Profsoyuznaya Str. 84/32, Moscow 117997, Russia}

\author{M.~LaRose}
\affiliation{Department of Physics and Astronomy, West Virginia University, Morgantown, WV 26506, USA}
\affiliation{Center for Gravitational Waves and Cosmology, 
West Virginia University, Chestnut Ridge Research 
Building, Morgantown, WV 26505}

\author[0000-0002-2034-2986]{L.~Levin}
\affiliation{Jodrell Bank Centre for Astrophysics, School of Physics and Astronomy, The University of Manchester, Manchester, M13 9PL, UK}

\author[0000-0001-5229-7430]{R.~S.~Lynch}
\affiliation{Green Bank Observatory, P.O. Box 2, Green Bank, WV 24494, USA}

\author[0000-0001-7697-7422]{M.~McLaughlin}
\affiliation{Department of Physics and Astronomy, West Virginia University, Morgantown, WV 26506, USA}
\affiliation{Center for Gravitational Waves and Cosmology, 
West Virginia University, Chestnut Ridge Research 
Building, Morgantown, WV 26505}

\author{M.~Mingyar}
\affiliation{Department of Physics and Astronomy, West Virginia University, Morgantown, WV 26506, USA}
\affiliation{Center for Gravitational Waves and Cosmology, 
West Virginia University, Chestnut Ridge Research 
Building, Morgantown, WV 26505}

\author[0000-0002-4187-4981]{H.~Al Noori}
\affiliation{Department of Physics, University of California, Santa Barbara, Santa Barabara, CA 93106, USA}

\author[0000-0001-5799-9714]{S.~M.~Ransom}
\affiliation{National Radio Astronomy Observatory, 520 Edgemont Road, Charlottesville, VA 23903, USA}

\author{M.~S.~E.~Roberts}
\affiliation{New York University Abu Dhabi, Abu Dhabi, UAE} 
\affiliation{Eureka Scientific, Inc., 2452 Delmer St., Suite 100, Oakland, CA 94602, USA}

\author{A.~Schmiedekamp}
\affiliation{Department of Physics, The Pennsylvania State University, Ogontz Campus, Abington, Pennsylvania 19001, USA}

\author{C.~Schmiedekamp}
\affiliation{Department of Physics, The Pennsylvania State University, Ogontz Campus, Abington, Pennsylvania 19001, USA}

\author[0000-0002-7778-2990]{X.~Siemens}
\affiliation{Center for Gravitation, Cosmology, and Astrophysics, 
Department of Physics, University of Wisconsin-Milwaukee, P.O. Box 413, Milwaukee, WI 53201, USA}

\author[0000-0001-9784-8670]{I.~Stairs}
\affiliation{Dept. of Physics and Astronomy, UBC, 6224 Agricultural Road, Vancouver, BC, V6T 1Z1 Canada}

\author[0000-0002-7261-594X]{K.~Stovall}
\affiliation{University of New Mexico, Albuquerque, NM 87131}

\author[0000-0002-9507-6985]{M.~Surnis}
\affiliation{Department of Physics and Astronomy, West Virginia University, Morgantown, WV 26506, USA}
\affiliation{Center for Gravitational Waves and Cosmology, 
West Virginia University, Chestnut Ridge Research 
Building, Morgantown, WV 26505}

\author[0000-0001-8503-6958]{J.~van Leeuwen}
\affiliation{ASTRON, the Netherlands Institute for Radio Astronomy, 
Oude Hoogeveensedijk 4, 7991 PD Dwingeloo, The Netherlands}
\affiliation{Anton Pannekoek Institute for Astronomy, 
University of Amsterdam, Postbus 94249, 1090 GE Amsterdam, The Netherlands}

\correspondingauthor{A.~E.~McEwen}
\email{aemcewen@uwm.edu}

\begin{abstract}
The Green Bank North Celestial Cap (GBNCC) pulsar survey will cover the entire northern sky ($\delta > -40\degr$) at 350\,MHz, and is one of the most uniform and sensitive all-sky pulsar surveys to date. We have created  a pipeline to re-analyze GBNCC survey data to take a 350\,MHz census of all pulsars detected by the survey, regardless of their discovery survey. Of the 1413 pulsars in the survey region, we were able to recover 670.  For these we present measured signal-to-noise ratios (S/N), flux densities, pulse widths, profiles, and where appropriate, refined measurements of dispersion measure (656 out of 670) and new or improved spectral indices (339 out of 670 total, 47 new, 292 improved). We also measure the period-pulse width relation at 350\,MHz to scale as $W \propto P^{-0.27}$. Detection scans for several hundred sources were reanalyzed in order to inspect pulsars' single pulse behavior and 223 were found to exhibit evidence of nulling. With a detailed analysis of measured and expected S/N values and the evolving radio frequency interference environment at 350\,MHz, we assess the GBNCC survey's sensitivity as a function of spin period, dispersion measure, and sky position. We find the sky-averaged limiting flux density of the survey to be 0.74\,mJy. Combining this analysis with PsrPopPy pulsar population simulations, we predict 60/5 non-recycled/millisecond pulsar discoveries in the survey's remaining 21,000 pointings, and we begin to place constraints on population model parameters.
\end{abstract}


\section{Introduction}
The Green Bank North Celestial Cap (GBNCC; \citealt{slr+14}) pulsar survey began in 2009 and, when complete, will cover the entire sky accessible to the 100\,m Robert C. Byrd Green Bank Telescope (GBT; $\delta\geq-40\degr$, or 85\% of the celestial sphere) at 350\,MHz. As of mid-2019, the survey is 85\% complete and 161 pulsars have been discovered, including 25 millisecond pulsars (MSPs) and 16 rotating radio transients \citep[RRATs;][]{mll+06}. Timing solutions for these discoveries have been published in \cite{slr+14}, \cite{kkl+15}, \cite{kmlk+18},  \cite{lsk+18}, and \citet{acd+19}, and more are forthcoming.  As such, this constitutes one of the largest and most uniform pulsar surveys to date. 

In addition to the newly discovered pulsars, the uniform coverage of GBNCC allows a robust re-assessment of the known pulsar population with reliable flux density measurements.  Here we present a detailed search for all known pulsars in the GBNCC footprint.
We find that 572 previously published pulsars and 98 unpublished pulsars have been re-detected by the survey pipeline and visually confirmed, comprising 670 detections in total, the largest low-frequency, single-survey sample.  Similar to previous efforts based on results from the Parkes Multibeam Pulsar Survey (PMPS) and the Pulsar Arecibo L-band Feed Array (PALFA) survey \citep[e.g., see][]{lfl+06,slm+14,lbh+15}, we conduct a detailed  analysis of the GBNCC pulsar survey and compare its sensitivity with that of other surveys in overlapping regions of sky. Flux densities at 350\,MHz ($S_{350}$) are presented for all detections, as well as 
pulse widths and profiles. 

In \S\ref{sec:methods}, we outline the process used to generate a comprehensive list of pulsars as well as predicting and measuring signal-to-noise ratios (S/N) of detections in the survey. In \S\ref{sec:results}, we present the recovered S/N and flux density measurements for all detected pulsars as well as measurements of pulse width, dispersion measure, and spectral index. We also present the profiles for all of these pulsars. In \S\ref{sec:sensitivity}, we discuss how the GBNCC survey is performing compared to expectations and RFI characteristics of the survey, and remark on interesting detections and notable non-detections. We also discuss the implications of our results for the Galactic pulsar population. Finally, in \S\ref{sec:conc}, we summarize the main conclusions of this analysis.

\section{Sample Assembly and Data Reduction}

\label{sec:methods}
The GBNCC data set as of late fall 2018 included $\sim$108,000 120\,s pointings, each tagged with a unique beam number. Each dual-polarization observation was taken with the GBT over the past $\simeq$10 years. The survey utilizes the GUPPI backend, with a sampling time of 82\,$\mu$s and 100\,MHz of bandwidth centered at 350\,MHz \citep[for more information on the observing setup for the GBNCC survey, see][]{slr+14}. We began by organizing a comprehensive list of all known pulsars with parameters that were available for use, whether they were published or not. By utilizing the Australia Telescope National Facility (ATNF) pulsar catalog\footnote{\url{http://www.atnf.csiro.au/research/pulsar/psrcat}} \citep[v1.59,][]{mhth05}, we amassed the bulk of the sources from the list of all published pulsars and their positions on the sky as well as their spin parameters and other relevant quantities (dispersion measure, etc.). Discovery parameters are also available for additional pulsars that have not been published but were detected in a number of other recent or ongoing surveys. Many of these surveys, including AODrift \citep{dsm+13}, SUPERB \citep[][Spiewak et al., 2019, in prep.]{kbj+18}, GBT 350\,MHz Drift \citep{blr+13}, PALFA \citep{cfl+06,lbh+15}, LOTAAS \citep{scb+19}, and HTRU-South \citep{kjs+10} include pulsars that are in GBNCC survey area, and so were included in the list. More information on these surveys is included in Table \ref{tbl:surv}. Furthermore, we included the list of pulsars that had been discovered in the search pipeline for the GBNCC survey. We then limited this list to pulsars within the range of the survey, i.e., pulsars with $\delta>-40\degr$.  In total, this list contained 2299 pulsars. We determined which pulsars were within 30\arcmin\ (FWHM of GBT at 350\,MHZ) of completed GBNCC pointings, adjusting when necessary to compensate for large ($>$30\arcmin) uncertainties in pulsar position. This reduced the total number of pulsar candidates to 1413. We could then match each pulsar with the GBNCC beams closest to its position before beginning to process the data. 

Radio frequency interference (RFI) excision is the first step of GBNCC data analysis, and is done primarily with the \texttt{rfifind} tool from the \texttt{PRESTO}\footnote{\url{http://www.cv.nrao.edu/~sransom/presto/}} pulsar data analysis software package \citep{smr+01} as described in \S3.1 of \citet{slr+14}. We also performed an analysis of the \texttt{rfifind} output files spanning the lifetime of the GBNCC survey up to late 2018 (roughly 83\% of the total survey) to characterize the effects of RFI over the course of the survey. These files contain information about which frequency channels were masked due to RFI for every 120 second scan in the survey. For a particular scan, the effective bandwidth $\Delta\nu$ is the total 100\,MHz bandwidth of the GBT 350\,MHz receiver multiplied by the ratio of unmasked to total channels for that scan, minus an additional 20\,MHz for rolloff. 

In some cases, the \texttt{rfifind} masks were insufficient to remove additional RFI that was either narrow in frequency space or brief in time. The latter often appears as a very bright burst at $\sim$0 DM for portions of the observation. To mitigate this, we employed some additional narrowband flagging in the \texttt{PRESTO} \texttt{prepfold} command as well as removing corrupted portions of the scan in the time-domain. Note that these changes also alter the values for $\tau_{\rm obs}$ and $\Delta\nu$ which consequently change the measured S/N for a given observation. For this reason, we calculate the fraction of data points from the observations that were not omitted in processing and multiply the total bandwidth by this fraction.  

After removal of RFI, we dedispersed and folded the observations at each pulsar's rotational period and integrated the profiles to obtain a single average profile for each observation. For the vast majority of sources included in this analysis, a precise ephemeris from the ATNF catalog was used to perform the folding. In all other cases, only the discovery parameters (period, DM and, if known, period derivative) were used. We also repeated this process while allowing dispersion measure to vary and, in some cases, also allowing variations in period and period derivative. This second iteration allows for fine-tuning previously published parameters at the cost of potentially finding bright RFI, which will often occur when attempting to detect low-DM pulsars as sources of RFI have DM = 0\,pc\,cm$^{-3}$. The 120 second observation times utilized in the GBNCC survey limit sensitivity to period refinement, so fitting for period was only used to increase the S/N of detections of pulsars for which only discovery parameters were used, and no further timing analysis was done as a part of this study. All folded data were visually inspected to determine likelihood of an actual detection. In cases where RFI still existed in the data, we removed high order ($>$5) polynomials from the off-pulse regions of the profile. With folded profiles, we calculated a measured signal-to-noise ratio (S/N) \citep{lk04},
\begin{equation}
{\rm S/N}_{\rm meas} = \sum_{i=0}^{N_{\rm bin}} \frac{p_i-\bar{p}_{\rm off}}{\sigma_{\rm off} \sqrt{W N_{\rm bin}/P}}~\gamma,
\label{eq:snm}
\end{equation}
where $N_{\rm bin}$ is the number of bins across the pulse profile, 
$p_i$ is the value of bin $i$, 
$\bar{p}_{\rm off}$ is the mean of the off-pulse bins, 
$\sigma_{\rm off}$ is the standard deviation of the off-pulse bins, $W$ is the on-pulse width in seconds, $P$ is the pulsar spin period in seconds, and $\gamma$ is a correction factor. When continuous signals are assigned to a finite number of bins in the profile during the folding process in \texttt{PRESTO}, their intensity is ``smeared" over the neighboring bins, resulting in correlations in the bins' intensities. This correction, dubbed $\gamma$, depends on the sampling time and the number of bins in the profile, which (for this study) is dependent on the pulsar spin period. Typical values are close to 0.95. The number of bins $N_{\rm bin}$ was determined by the pulsar period as follows: profiles for pulsars with periods shorter than 1.7\,ms had 28 bins, periods shorter than 10\,ms had 50 bins, periods shorter than 50\,ms had 128 bins, and all others had 200 bins. This prescription retains sensitivity to long-period pulsars but avoids bin widths corresponding to time intervals smaller than the sampling time of 82\,$\mu$s. Pulse widths were determined with a standard process. First, sigma-clipping was used to find the off-pulse region. Then, the peak value above the noise floor was identified, and bins on either side of the peak were added to the on-pulse width. This process was repeated, adding bins on the sides of the peak until we reached bins within 2$\sigma$ of the mean of the noise. The edges of the pulse were found by fitting lines to the two bins on either side of the pulse and finding the fraction of the outermost bins that were above the noise floor. At this point, we consider the full on-pulse width to be determined. Each profile was then checked by eye, and corrections to the on-pulse region were made. Any components of the pulse width that were distinct from the main pulse were determined using the same algorithm. To determine the sensitivity of uncertainties in S/N from the choice of the number of on-pulse bins, noisy Gaussian pulses were simulated and various width choices were used to measure the fractional error on S/N. From this test, it was found that on-pulse widths that exceed at least one $\sigma$ beyond the Gaussian mean were sufficient to greatly reduce the fractional uncertainty on S/N. Beyond this, adding bins had little effect on this fractional uncertainty - so, pulse widths were chosen to encompass all of the pulse visible above the noise.  In some cases, additional RFI features were removed prior to the determination of $W$ to minimize errors in $W$ and S/N (see \S\ref{sec:rfi}).  

Characteristic measurements of pulse width include measurements at both 50\% and 10\% of the pulse profile's maximum amplitude (hereafter $W_{50}$ and $W_{10}$, respectively). These widths are dependent on both pulse period and observing frequency, so measurements at 350\,MHz help to fill out the low-frequency regime for a wide range of pulse periods. However, the noise floor in some pulsars limits the ability to determine $W_{10}$ robustly. Note also that $W_{50}$ and $W_{10}$ are distinct from $W$, which includes all bins that contain the pulse signal, and so $W$ is generally slightly larger than $W_{10}$.

The expected S/N of a pulsar  can be estimated as \citep{dtw+85,lk04} 
\begin{equation}
{\rm S/N}_{\rm exp} = \frac{ S_{350}G\sqrt{N_{\rm pol}\tau_{\rm obs}\Delta\nu}} {T_{\rm sys}\beta} \sqrt{\frac{P-W}{W}} f(\theta),\label{eq:sne}
\end{equation}
where $S_{350}$ is the flux density at 350\,MHz, 
$G=2$\,K/Jy is the gain of the Green Bank Telescope \citep{slr+14}, 
$N_{\rm pol}=2$ is the number of polarizations recorded, 
$\tau_{\rm obs}=120$\,s is the length of the observation, 
$\Delta\nu$ is the bandwidth in MHz after removing RFI (see \S\ref{sec:rfi}), 
$T_{\rm sys}$ is the system temperature (including the sky temperature at the source position, receiver temperature $\simeq$ 20\,K, and CMB temperature $\simeq$ 3\,K), 
$\beta\simeq1.1$ is an instrument-dependent correction factor due to downsampling the data to 2 bits \citep{lk04}, 
and $f(\theta)$ is a radial Gaussian factor accounting for sensitivity degradation as a function of angular offset from the center of the circular beam $\theta$. The sky temperature in the direction of each pulsar was determined by using the measurements made by \cite{hks+81} for the beam positions, scaled to 350\,MHz using with the spectral index therein, $-$2.6.

Where possible we use flux densities at other frequencies and previous measurements of spectral index ($\alpha$, with $S_\nu \propto \nu^{\alpha}$) from the ATNF catalog to determine an expected flux density at 350\,MHz and the expected S/N \citep{mhth05}. In cases where there was no published value for $\alpha$ but flux densities at both 400\,MHz and 1400\,MHz were published, we determine a spectral index using a simple power law. In all other cases, we assume a spectral index of $-1.4$ \citep{blr+14} to estimate the flux density at 350\,MHz. We also calculate the measured flux density of each pulsar by inverting Equation \ref{eq:sne} and using measured values for S/N (determined from Equation \ref{eq:snm}) and pulse width. Comparing the expected flux density to our measurements can both roughly confirm our current models for pulsar emission as well as aid in explaining non-detections.


\section{Pulsar Flux Density Census at 350\,MHz}
\label{sec:results}

\begin{figure*}[ht]
\centering
\plotone{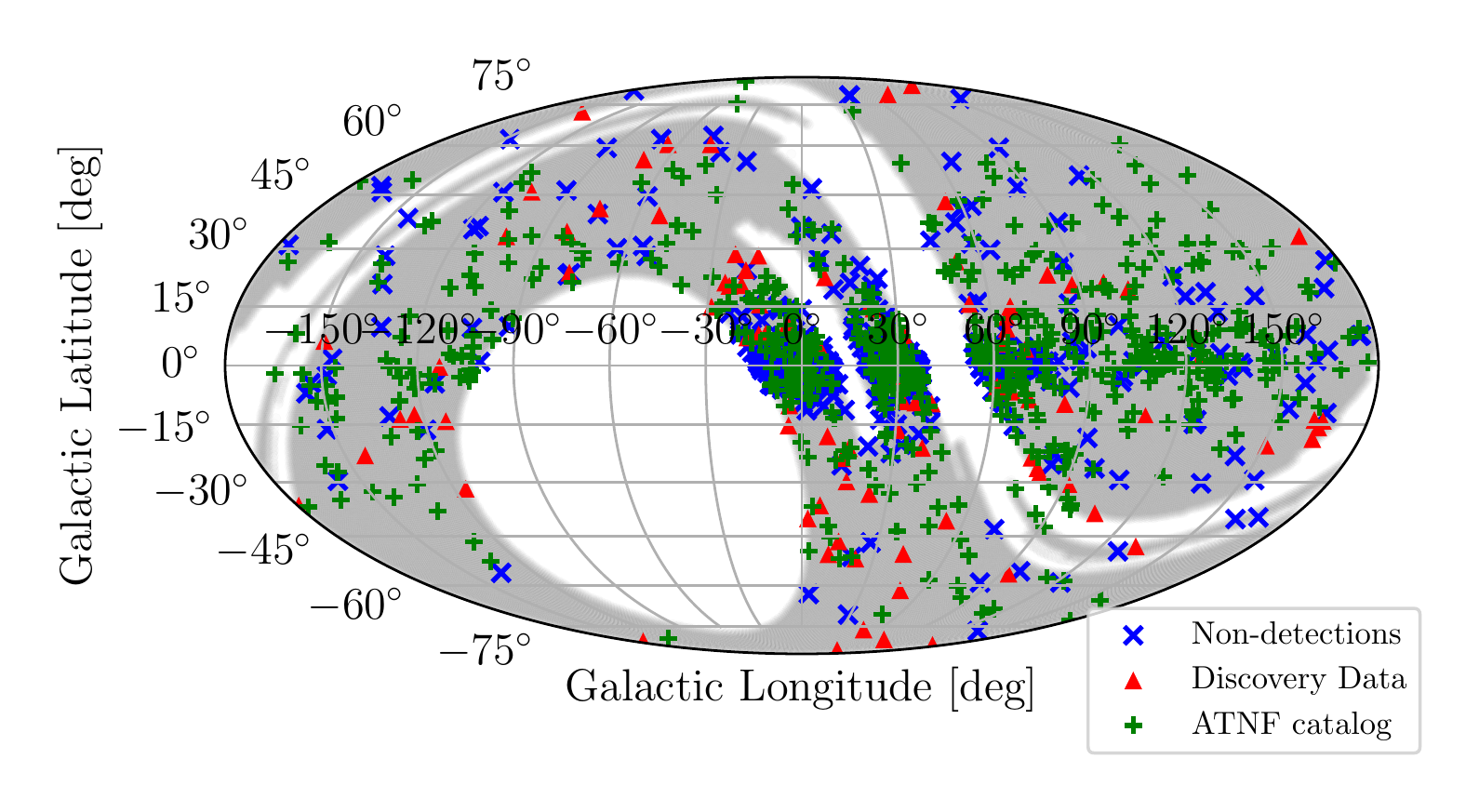}
\caption{Sky map with pulsars from overlapping surveys, plotted in Galactic coordinates as a Mollweide projection. The shaded regions indicate completed GBNCC observations. Detected pulsars from the ATNF catalog and pulsars that were detected using discovery parameters from overlapping surveys are differentiated by marker type, with green plus symbols indicating pulsars from the catalog and red triangles indicating pulsars from the surveys listed in Table \ref{tbl:surv}. Pulsars that were not detected are plotted as blue ``x" symbols.}\label{fig:skymap}
\end{figure*}

We detected 670 pulsars out of a total of 1413 in the survey area, and these detections are listed in Table \ref{tbl:det} in Appendix 1. For all following analysis, the beams corresponding to the brightest detections (highest S/N) were used, as these are most likely to represent the pulsars' flux density. Along with pulsar names, we provide several relevant quantities: dispersion measure from searching with \texttt{PRESTO} \citep{smr+01}, MJD of the brightest detection, angular offset from the center of the beam, $W_{50}$, $W_{10}$ (when S/N was large enough), detection S/N, 350\,MHz flux density measured from the GBNCC data, and measured spectral index $\alpha$ (see \S\ref{sec:spindx}). Uncertainties on the S/N and flux densities were calculated using standard error propagation from equations \ref{eq:snm} and \ref{eq:sne} and uncertainties on bandwidth, temperature, and $\theta$ of 5\,MHz, 10\,K, and 0.5 degrees, respectively. Among these are 66 millisecond pulsars (MSP), defined here as pulsars with spin periods shorter than 30\,ms. The integrated pulse profiles for all of the brightest detections are shown in Figures \ref{fig:prof0} $-$ \ref{fig:prof11} along with pulsar names, dispersion measure, and flux density. Figure \ref{fig:skymap} shows all detected pulsars plotted by their Galactic positions, and different markers indicate whether or not the pulsars were from the ATNF catalog or were a part of one of the other survey lists mentioned above.

\subsection{Comparison Between the GBNCC and Overlapping Pulsar Surveys}

\begin{deluxetable*}{lcccc}
      \tablecaption{Pulsar Survey Comparison \label{tbl:surv}} 
      \tablehead{
\colhead{Survey} & \colhead{Central Frequency} & \colhead{Limiting Flux Density\tablenotemark{a}} & \colhead{Detections\tablenotemark{b}} & \colhead{Reference}  \\ 
& \colhead{(MHz)} & \colhead{(mJy)} & & }
\startdata
AODrift \dotfill & \phn327 & 0.59  & 7/13 & \cite{dsm+13}\\
HTRU$-$S (low-lat) \dotfill & 1352 & 0.40   & 0/9\phn & \cite{kjs+10}\\
HTRU$-$S (med-lat) \dotfill & 1352 &  0.95   & 3/27 & \cite{kjs+10}\\
HTRU$-$S (high-lat) \dotfill & 1352 & 1.2\phn   & 1/8\phn & \cite{kjs+10}\\
SUPERB \dotfill & 1352 & 0.4\phn  & 2/15  & \cite{kbj+18}, Spiewak et al. (in prep)\\
LOTAAS  \dotfill& \phn134 & 0.63 & 10/39\phn & \cite{scb+19} \\
PALFA  \dotfill& 1400 & 0.23 & 0/29 & \cite{lbh+15}  \\
GBT350 \dotfill & \phn350 & 0.59  & 3/6\phn & \cite{blr+13} \\
GBNCC \dotfill & \phn350 & 0.70  & 72/72\phn & \cite{slr+14} \\
\enddata
\tablenotetext{a}{Averaged over the survey area and scaled to 350\,MHz.}
\tablenotetext{b}{Number of detections of pulsars from this survey by GBNCC/number of pulsars from this survey within the GBNCC survey area.}
\tablecomments{Information about individual detections is reported in Table \ref{tbl:det}.}
\end{deluxetable*}

Out of the 210 pulsars with discovery parameters that are not currently listed in the ATNF catalog, 98 were detected. Names, central frequencies, scaled limiting flux densities, and the ratio of detected to processed pulsars are given for each survey in Table \ref{tbl:surv}. It should be noted that there are many pulsars from these surveys (excluding GBNCC) in regions of the sky where the GBNCC survey has yet to observe, and so they may be detected in the future; these pulsars are not included in the counts listed in Table \ref{tbl:surv}. Three of these surveys (SUPERB, HTRU-S, PALFA) were conducted at higher frequencies,, where average sky temperature (especially near the Galactic plane) is much lower. This reason and the increased sensitivity to high DM pulsars at high frequency is useful for diagnosing missed detections. Because these pulsars have neither published flux densities nor spectral indices, reasons for missed detections cannot be determined more robustly than those due to sky temperature, position relative to the survey, extreme nulling/intensity variation, and high DM/short periods. It is also possible that for some of these pulsars, the discovery parameters may not be precise enough to be found in this analysis.

The most surprising missed detections come from the GBT350, AODrift, and LOTAAS surveys, which all have comparable sensitivities and frequencies. In an effort to explain why these pulsars were missed, all of the discovery plots were checked against our results, and acceleration searches were run. Three pulsars (J0100+69 and J0121+14 from LOTAAS, and J1854+36 from AODrift) that were originally missed were found on the second trial, as the DM used in the first run was not close enough to the DM at which the pulsar was discovered. For the majority of pulsars that were not detected after re-running the pipeline, the discoveries were quite dim. The LOTAAS survey also has much longer integration times (60 minutes), which significantly improves the chances of the survey detecting pulsars which may be faint and/or nulling. When checking the discovery plots, it became clear that both of these effects were common to many of the missed pulsars. Some pulsars even appeared to exhibit nulling with `off' times as large as 100 seconds. Nulling behavior was also seen in many cases for the AODrift survey. For the GBT350 missed pulsars, all three of those that were missed were faint, and several GBNCC beams in which the pulsars were most likely to be found had RFI that spanned the entire 100\,MHz band.

Eight binary pulsars that were originally discovered in the GBNCC survey were not detected in the first pass of this pipeline. These pulsars required acceleration searches, which are automatically performed as a part of the search pipeline, but not here. As a part of the missed pulsar analysis, we ran an additional acceleration search using \texttt{ACCELSEARCH} from within the \texttt{PRESTO} package, and they were all detected. We also reprocessed data for 15 binary pulsars from the ATNF catalog with short ($\leq$ 0.5 day) orbital periods that were not detected in the first pass using acceleration searches; none of these were detected.

Pulsars with long periods (greater than 2.5\,s) were also followed up with a search for single pulses. Because these pulsars would only be observed for at most 48 pulses, non-detections are more common. To address this, we implemented \texttt{single\_pulse\_search.py} from the \texttt{PRESTO} package, which searches a range of dispersion measures to find bright single pulses in the data and characterize them by their S/N. In this way, a pulsar that is not detected via a periodicity search may be found by individual pulses. However, we were still unable to find these pulsars using this method.

\subsection{Spectral Indices}
\label{sec:spindx}
\begin{deluxetable}{cccc}
\tablecaption{Broken Powerlaw Spectral Indices \label{tbl:spindx}}
\tablehead{\colhead{PSR} & \colhead{$\alpha_{l}$\tablenotemark{a}} & \colhead{$\alpha_{h}$\tablenotemark{a}} & \colhead{Break Frequency} \\
 & & & (MHz)}
\startdata
J0034-0534 & 0.6(3) & $-$3.1(2) & 181 \\ 
J0218+4232 & 1.15(7) & $-$2.7(4) & 149 \\ 
J1900-2600 & 0.2(4) & $-$1.89(15) & 204 \\ 
J2002+4050 & 0.2(16) & $-$1.51(18) & 378 \\
\enddata
\tablenotetext{a}{Spectral indices below ($\alpha_l$) and above ($\alpha_h$) the break.}
\tablecomments{Quantities in parentheses are uncertainties in the last digit.  See Figure~\ref{fig:spindx} for the corresponding plots.}
\end{deluxetable}

\begin{figure*}[ht]
\epsscale{1.}
\plotone{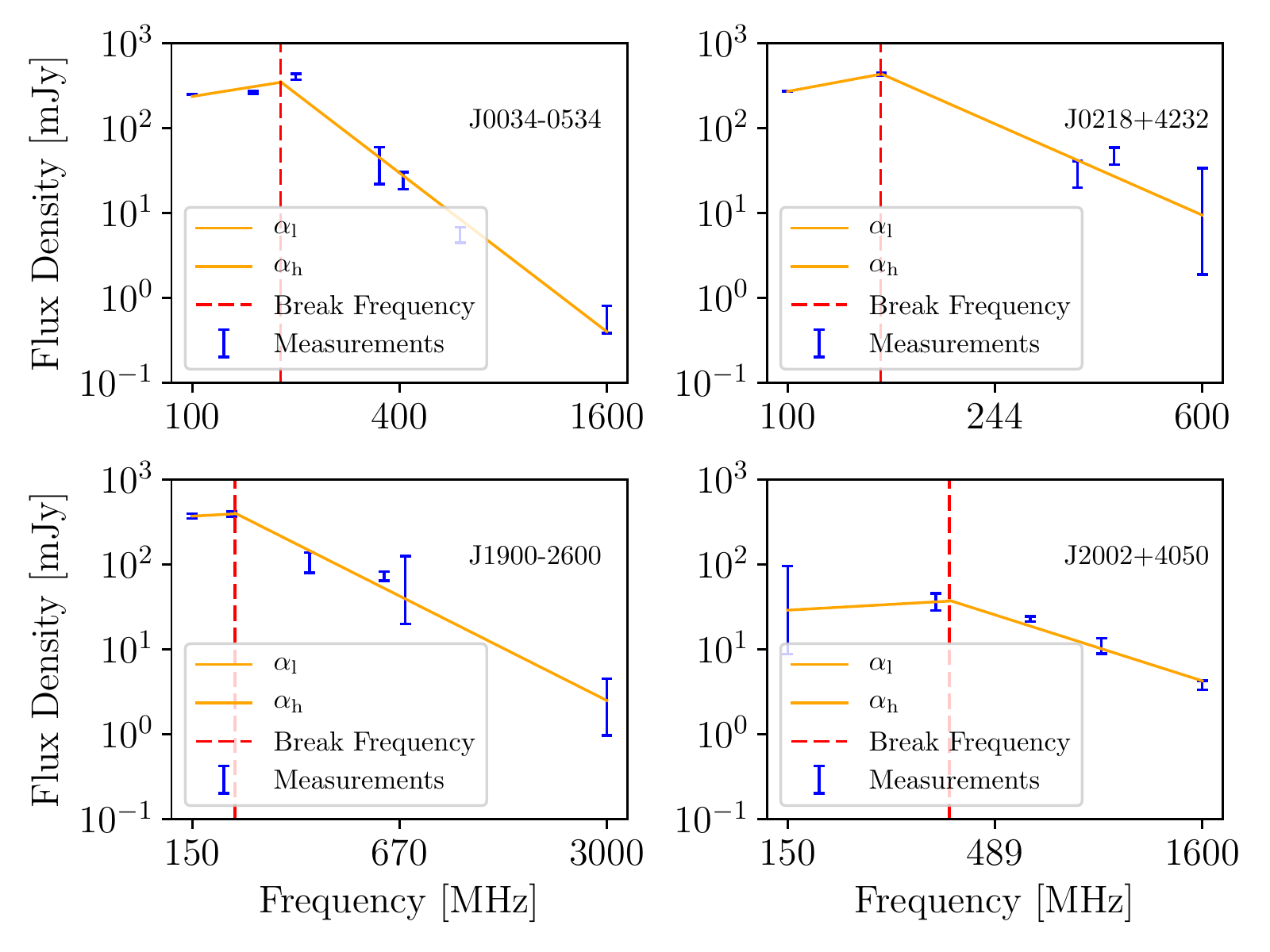}
    \caption{Pulsars with broken power-law spectral indices. We plot all available measurements of flux density in the ATNF catalog as well as the 350\,MHz measurements made in this study against observing frequency. We  fit two disjoint lines to the low- and high-frequency measurements (orange solid lines). The red dashed line indicates the frequency of the turnover in the spectrum, determined by finding the point at which the two lines match up. Information for these measurements is presented in Table \ref{tbl:spindx}.}
    \label{fig:spindx}
\end{figure*}

Many previously published spectral indices were determined from flux measurements from high-frequency surveys \citep[e.g., see][]{jsk+18}. Therefore, low frequency surveys like the GBNCC survey provide more stringent constraints on these calculations.  Results from this analysis are listed in Table \ref{tbl:det}. 
The majority of the pulsars in this data set follow a single power law, or do not have enough ($>$2) flux density measurements to fit multiple power law functions.  However, there are a small number of cases where the emission is better fit by a broken power law, defined instead as a piecewise function composed of two power laws. 
All 339 pulsars for which we measured spectral index had three or more flux measurements (including our 350\,MHz measurements) and were checked by eye to determine whether or not a broken line fit was appropriate. Four pulsars fit these criteria. For these pulsars, we fit two lines, one for high frequency flux density measurements and one for low frequency. The breaking point for the power law was determined by finding the maximum change in the derivative of flux density with respect to frequency. A similar analysis was done in \cite{mkb+17}. Plots of these cases are provided in Figure \ref{fig:spindx} with both indices included. These plots also display the best-fit line to all measured flux densities. The measured values of $\alpha_{l}$ and $\alpha_{h}$ are reported in Table \ref{tbl:spindx}.

\subsection{Comparison of Dispersion Measure with Catalog Values}

The relatively low frequency of the GBNCC survey allows much higher precision DM measurements than typical 1400-MHz surveys, as dispersion across the band scales as $\nu^{-2}$. As pulses propagate through the interstellar medium, this dispersion results in a frequency dependent delay that smears out the arrival time of the pulse. Tools within the \texttt{PRESTO} package adjust for this, shifting the low frequency portion of the signal back in time to line up the pulse across the band. Using the \texttt{dmsearch} flag contained within the \texttt{PRESTO} command \texttt{prepfold}, we processed each of the pulsars and recovered more accurate values of DM. The program adjusts for dispersion and then folds the data at the pulsar's period to line up the pulses in both time and frequency. When \texttt{dmsearch} is off, the program does not tune the DM to maximize S/N; otherwise, the DM which aligns the pulses in frequency is returned as a new DM. In some cases, RFI caused the DM searching algorithm to return erroneous values for DM, and so we were unable to refine the dispersion measure. For these pulsars, we include the previously published DM in Table \ref{tbl:det} and mark them with a double dagger. More often, we were able to improve upon the previously published values of DM. Most of the discrepancies were small, but in some cases, our more precise DM measurement differed from the previous value significantly. For the pulsars with significant changes to their previously catalogued DM, we followed up with \texttt{TEMPO}\footnote{\url{http://tempo.sourceforge.net}} (maintained and distributed by Princeton University and the Australia Telescope National Facility). We split each detection into four subbands and created precise pulse times-of-arrival (TOAs) which can then be utilized to fit for DM. This method provides marginally more precise measurements, and so was only performed on pulsars with significant changes to previous DM measurements ($\geq$3$\sigma$). All newly measured DMs are presented in Table \ref{tbl:det}, and Table \ref{tbl:dm} highlights the pulsars which were followed up with \texttt{TEMPO} timing.

\section{Survey Sensitivity}
\label{sec:sensitivity}

\begin{figure*}
\plotone{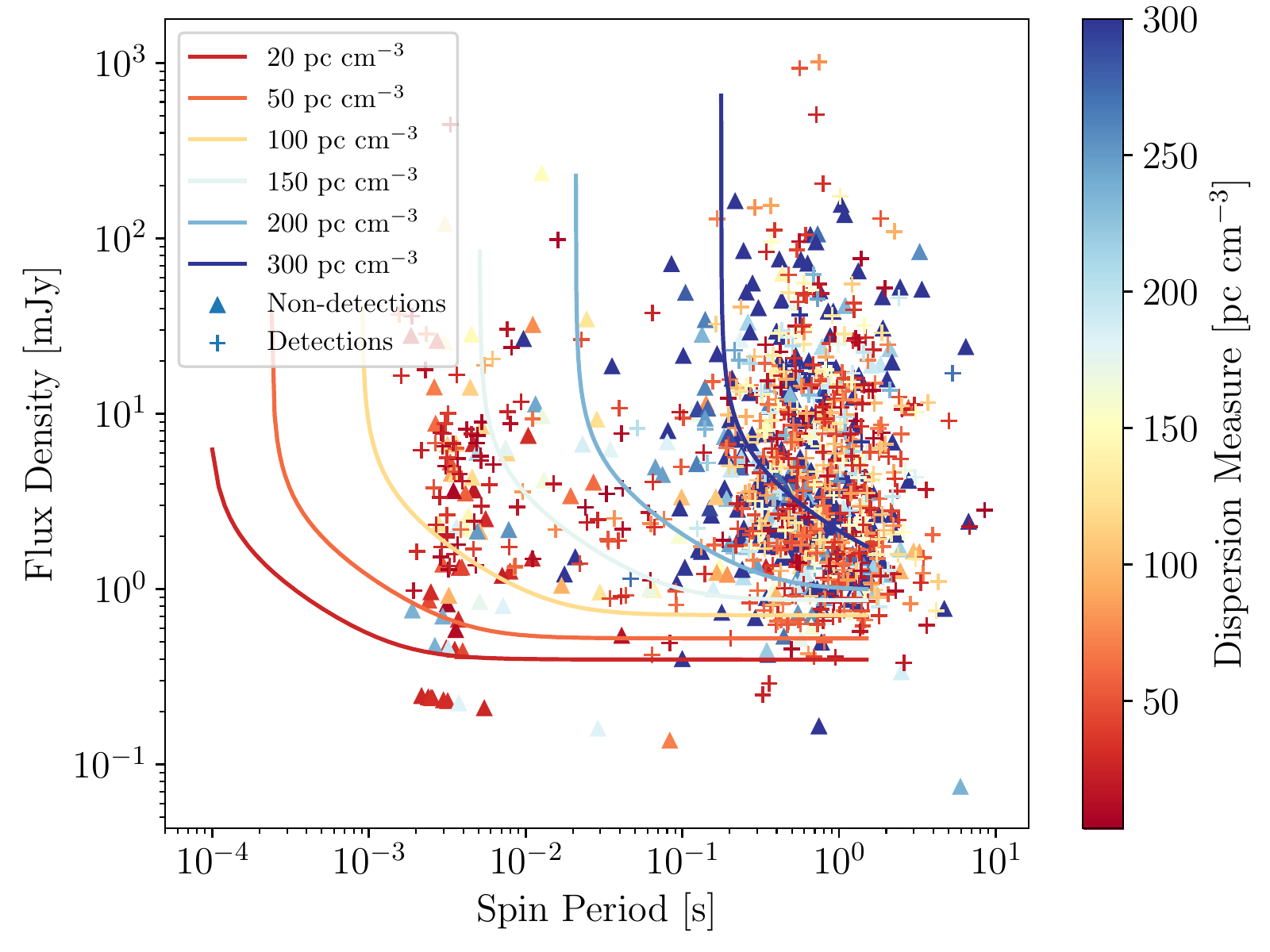}
\caption{Flux density sensitivity in the GBNCC as a function of pulse period. Assuming a duty cycle of 6\% and an average unmasked bandwidth of 67\,MHz (which incorporates a 20\,MHz rolloff in the bandpass), we plot the predicted lower limit on the flux density of detectable pulsars for dispersion measures of 20, 50, 100, 150, 200, and 300\,pc\,cm$^{-3}$. To determine the sky temperature for the curves, we found the average sky temperature as a function of DM using the sky temperatures at the positions of all detected pulsars. We then drew from this function the temperatures at each DM for which a curve is plotted. For the above DMs, the function returns 95, 126, 171, 208, 237, and 273\,K. We glean the minimum detectable S/N for the survey by matching the curves to the faintest detection. This was found to be $\sim$3.8. Higher DM pulsars are more susceptible to smearing, and so the likelihood of detection is decreased for high DM, short period pulsars. We also plot both the detections (plus symbols) and non-detections (triangles), which are colored by their DM.}\label{fig:senscurve}
\end{figure*}

\begin{figure}
\epsscale{1.3}
\plotone{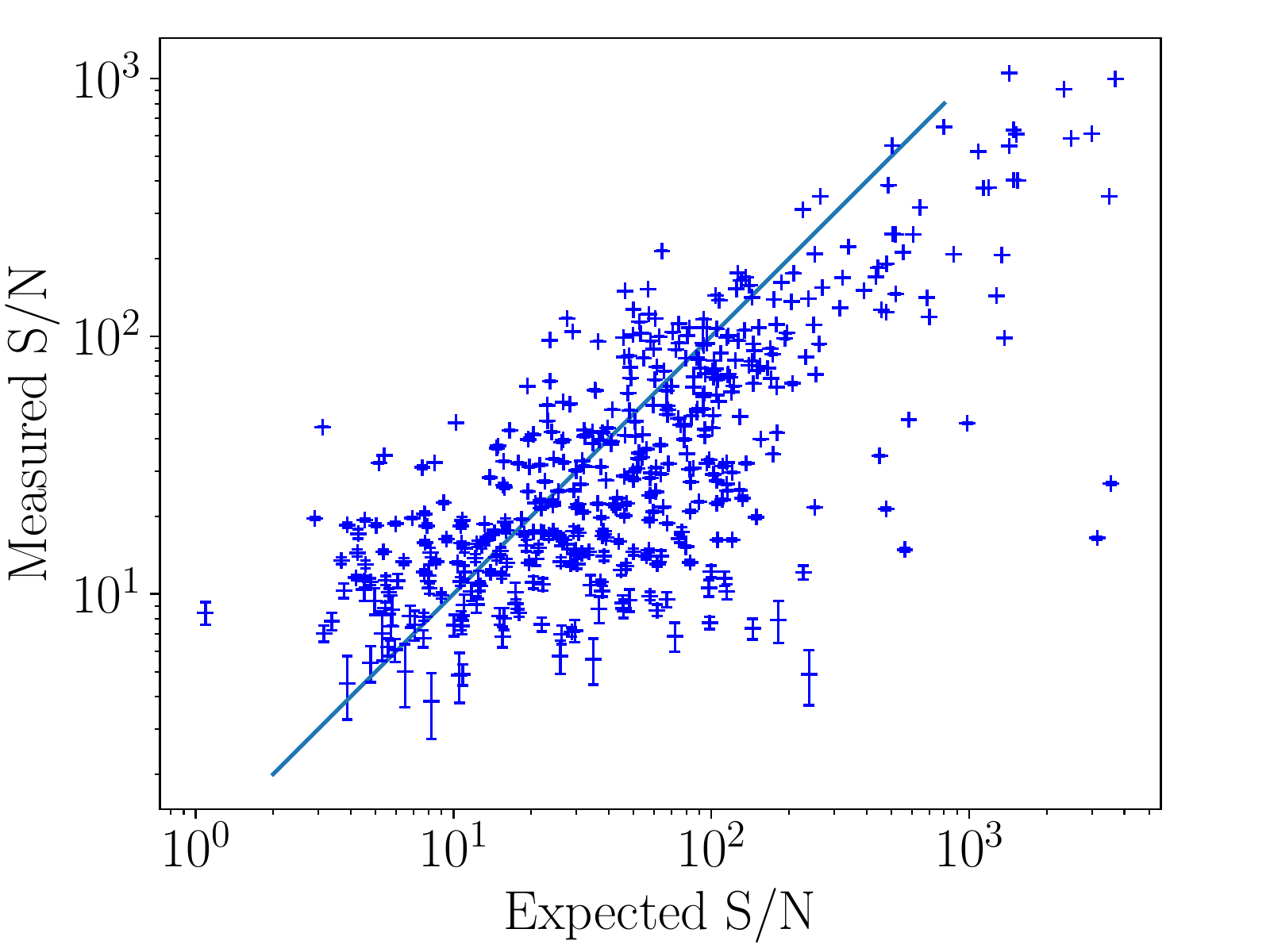}
\caption{Measured S/N vs.\ expected S/N for detections in the GBNCC survey. Extrinsic contributions to expected S/N include system temperature, telescope gain, scintillation, and offset from the beam center (newer pulsars without full timing solutions may have significant uncertainties in position). Errors in these quantities, previous flux measurements, and spectral indices increase the spread about unity, as does variable pulsar emission, i.e. nulling. }\label{fig:sn_mve}
\end{figure}

\subsection{Efficiency of GBNCC Survey}\label{subsec:efficiency}
In total, there were 5633 unique beams analyzed, yielding 1328 unique detections of the 670 pulsars. Given that there were 102948 beams that had been observed at the beginning of this project, this corresponds to an average number of detections per beam of $\sim$\,0.013 (0.063 detections per square degree), and $\sim$\,0.38 detections per hour of observing.
The ability to detect pulsars at 350\,MHz is limited most stringently by sky temperature and scattering in the interstellar medium (which correlates with dispersion). The expected S/N for detections is inversely proportional to system temperature, which is dominated by sky temperature near the Galactic plane. At 350\,MHz, this effect is quite significant, with temperatures approaching 1000\,K in this region. Scattering is especially detrimental in the detection of pulsars with short periods, as even a few milliseconds of smearing can eliminate the pulse entirely. Given a particular spin period and the estimated DM smearing, we can estimate the minimum flux density that will be detected by the survey. This relationship comes from solving Equation \ref{eq:sne} for flux density and assuming both an average sky temperature and duty cycle for the pulsars in the survey. Plotted in Figure \ref{fig:senscurve} are curves corresponding to a number of trial values of DM, showing the  sensitivity floor at those values. Because DM and sky temperature are correlated, we determined the average sky temperature for each curve that is plotted, resulting in an increase in minimum detectable signals for higher DM pulsars. Also plotted are flux density measurements for detections made by this survey and expected flux density measurements for the pulsars which were not successfully detected. The colors in the plot correspond to the dispersion measure of each pulsar, showing how pulsars that may be intrinsically bright enough to be detected can still be missed because of dispersive smearing and/or scattering. The minimum flux density expected to be measured in the survey (regardless of spin period) can be determined to be the asymptotic value of the DM curve corresponding to the faintest detection. This value is directly proportional to the minimum S/N which results in a detection, hereafter ${\rm S/N}_{\rm cut}$, which was found to be $\sim$\,3.8. For all detections, we plot both the expected S/N at 350\, MHz as well as the measured S/N of the detection. These are plotted in Figure \ref{fig:sn_mve} along with a line marking unity. There is a large spread about this line, due mostly to stochastic noise sources in the data (telescope noise, temperature fluctuations, scintillation, and variable pulsar emission). When examining these results, several of the more significant outliers were analyzed in closer detail. One of the three significant outliers in the lower right portion of the plot was found to be a new nulling candidate, and the other two were initially labeled as possible nullers that could not be verified without higher resolution observations.

\begin{figure}[ht]
    \centering
    \epsscale{1.2}
    \plotone{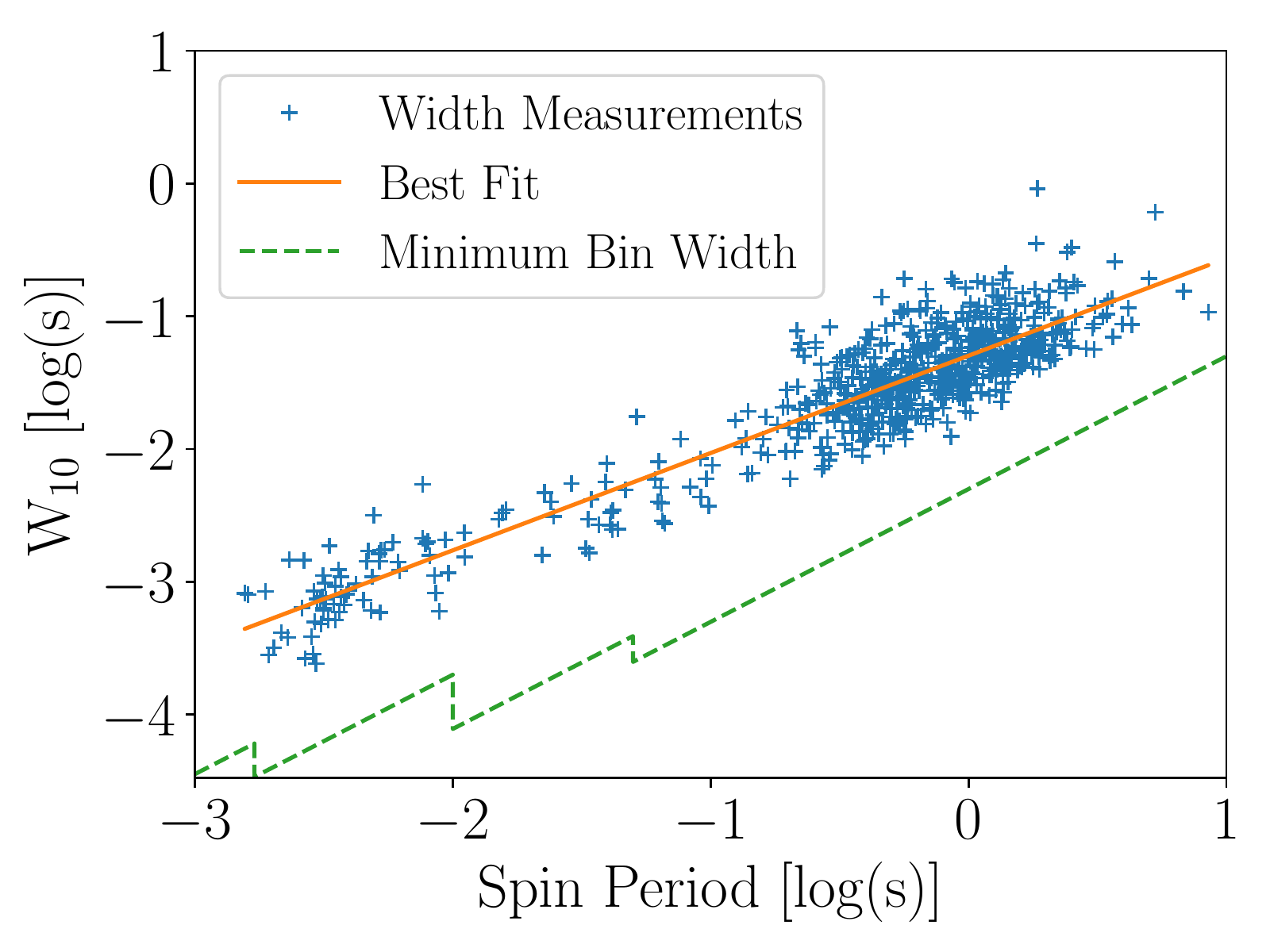}
    \caption{Pulse width at 10\% of the pulse maximum as a function of spin period. The solid line shows the line of best fit through the data, described by $W_{10}$ = $18.5\degr(4)P^{0.270(10)}$. The dashed line shows the minimum bin width as a function of period, as described in \S \ref{sec:methods}. }
    \label{fig:pvw}
\end{figure}

Low frequency observations can result in significant deterioration of the pulse due to scattering and scintillation effects, as residual dispersive time delay within a frequency channel with finite width increases as $\nu^{-3}$ and scattering roughly as $\nu^{-4}$ \citep{lk04}. Both of these phenomena result in a broadened pulse and subsequently a reduction in S/N.
To shed light on the causes for some of the missed pulsars, we calculate the expected S/N using information from both the catalog and information about the beams in which we expect to detect them. We predict flux density at 350\,MHz calculated as described in \S 2, determine the masked fraction of the closest beam to the pulsar's position (when measured), and determine $T_{sys}$ for the corresponding sky position. To determine $W$, we fit a line to our measurements of $W_{10}$ as a function of spin period and draw from this function. This allows for a measurement of the spin period-pulse width relation at 350\,MHz, supplementing previous measurements at other frequencies. This best fit line was measured to be $W_{10}$ = $18.5\degr(4)P^{0.270(10)}$, which is consistent with the relation determined in \cite{jk19} modulo a frequency-dependent scaling factor \citep[for a more in-depth analysis, see][]{cw14}.  This fit is shown in Figure \ref{fig:pvw}.

\begin{figure*}[ht]
    \epsscale{1.2}
    \centering
    \plotone{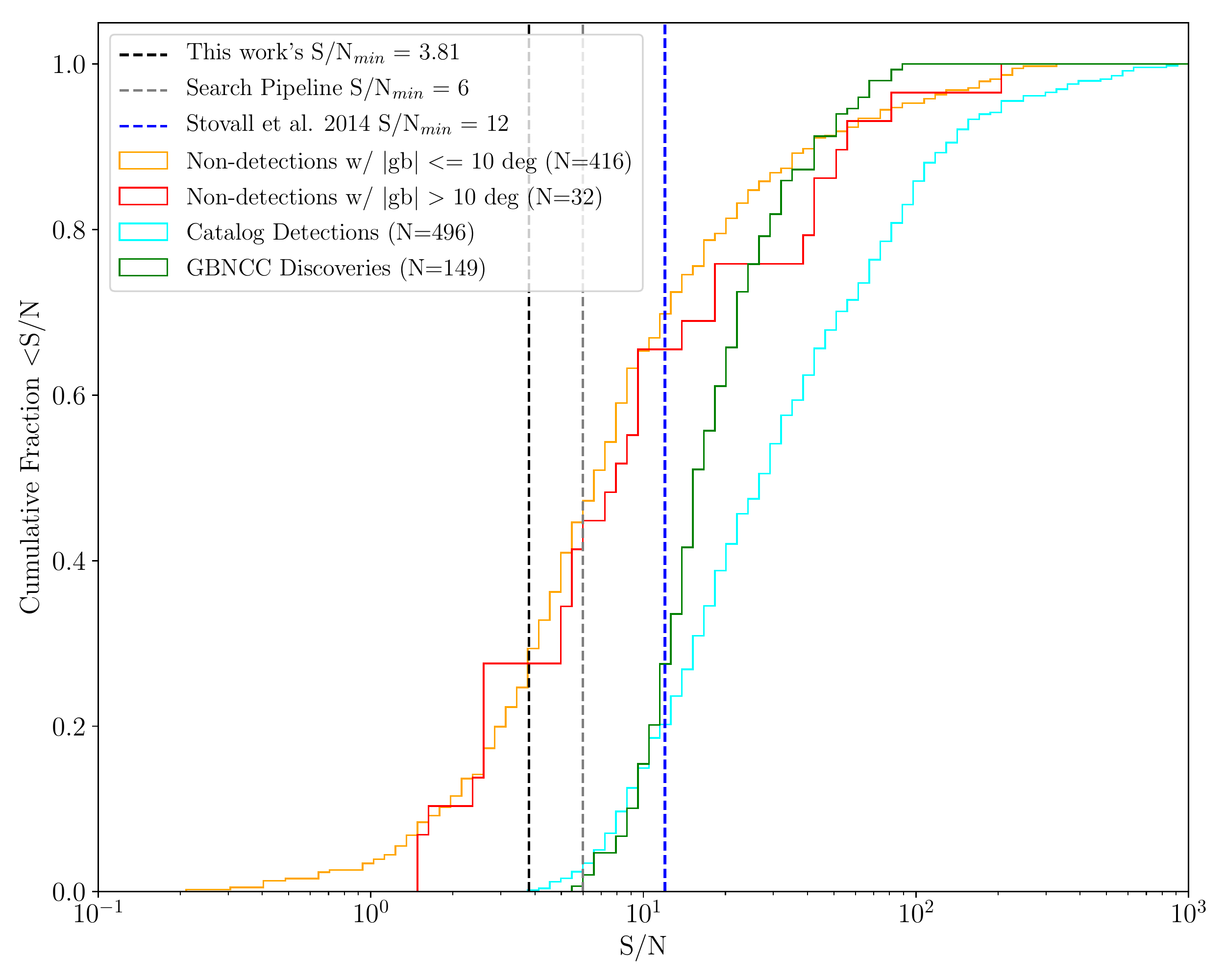}
    \caption{Histograms of measured S/N for detections and expected S/N for non-detections. Detections are differentiated by GBNCC discovery/catalog pulsars (green/cyan lines), and non-detections by distance from the Galactic plane (the red line indicates pulsars that are within 10$\degr$ from the plane, and the orange line indicates pulsars outside of this region). The dashed lines indicate three different S/N cutoffs: the first line, in black, show the minimum S/N detected in the survey; the second, in grey, indicates the significance down to which candidates are folded in the GBNCC search pipeline; and the third, in blue, shows the predicted S/N limit used in \cite{slr+14} to predict sensitivity of the survey.}
    \label{fig:sn_hist}
\end{figure*}

After drawing widths from either the catalog or the above function (based on availability of previous measurements of $W_{10}$ near 350\,MHz), we determined the expected S/N for all non-detections. These are plotted along with the measured S/N for all of the detections in Figure \ref{fig:sn_hist}. The detections have been divided between those found from the catalog and those discovered by the GBNCC survey, and non-detections are divided based on Galactic latitude. These divisions allow for direct comparison between the survey's ability to detect pulsars blindly as well as the limits placed on the survey by high temperatures and scattering near the Galactic plane. Included in the plot are three different S/N cutoffs placed during different stages of the survey. The least stringent cutoff of S/N = 12 comes from \cite{slr+14}, where it was used as an estimated cutoff for detection to predict the survey's sensitivity. At this S/N, $\simeq$75\% of non-detections are not expected to be detected. Pulsars close to the plane generally have lower S/N as the temperature is so high, while pulsars outside of the plane generally have smaller DM and temperature but more scintillation. The two detection curves show that the GBNCC is sensitive to intrinsically fainter pulsars, as the histogram is skewed toward lower measured S/N than those from the catalog. Note that there was one pulsar discovered by the GBNCC search pipeline with S/N = 5.98, which is the bin to the left of the search S/N cutoff.

\begin{figure*}[ht]
\epsscale{1}
\plotone{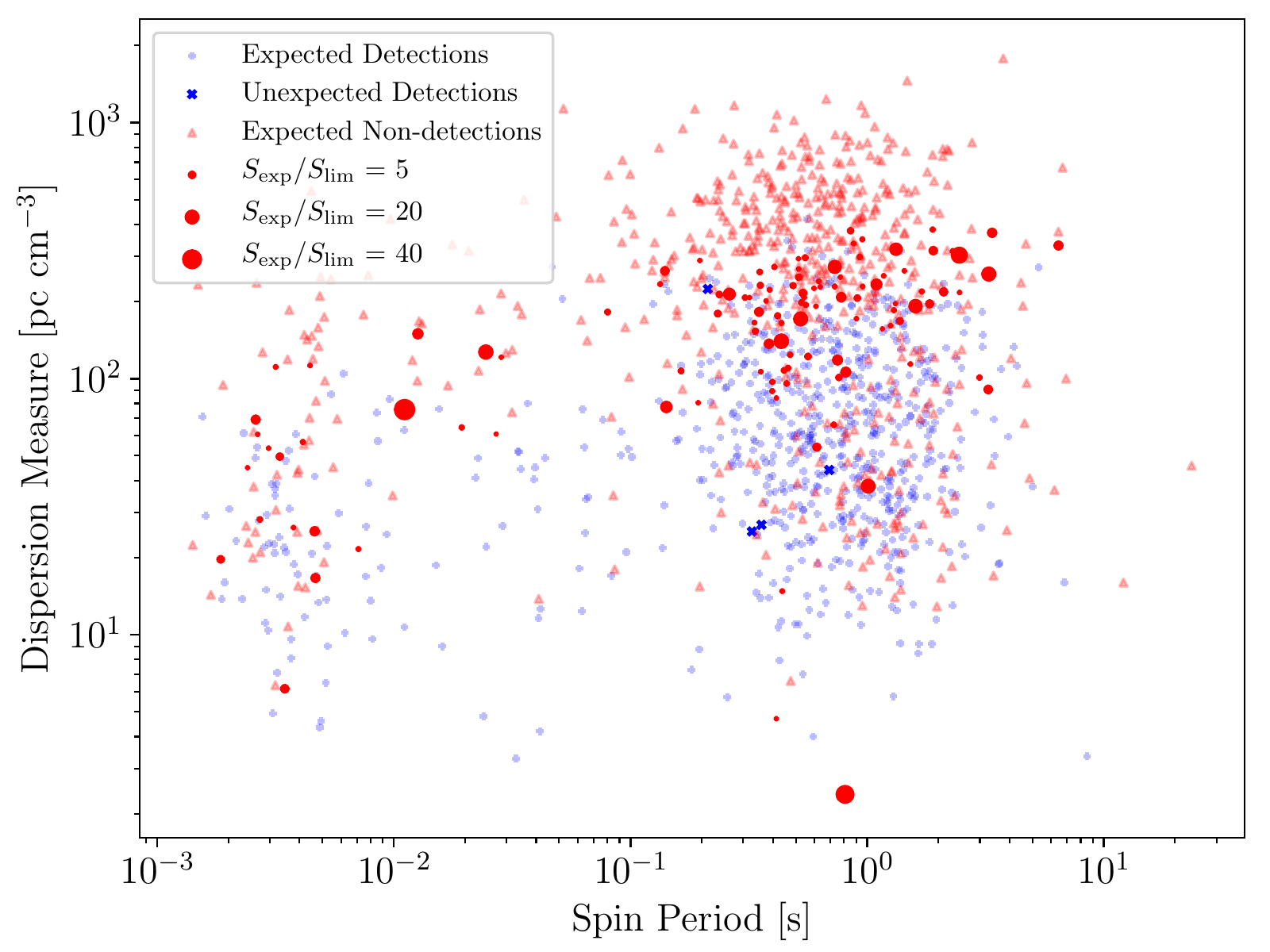}
    \caption{Period vs.\ DM for all included pulsars. Blue symbols indicate detections made by the survey, and red symbols indicate non-detections. Red triangles indicate missed pulsars that were not expected to be detected, in that they lie below the expected sensitivity of the survey. Red circles indicate missed pulsars that lie above their expected sensitivity, and so were unexpected non-detections (see \S\ref{subsec:efficiency} for details). Blue circles indicate detections that were expected, and blue x symbols indicate detection of pulsars with expected flux densities that were below our sensitivity limit. The area of these points is given by the ratio of expected flux density to the limiting flux density at the pulsar's position.}
    \label{fig:p0vsdm}
\end{figure*}

In Figure \ref{fig:p0vsdm}, we plot all pulsars' periods against their dispersion measure. Each point's color and shape describe whether or not the pulsar was detected, and if not, whether we expect to have detected it. Missed detections that were unexpected are plotted with point sizes reflecting the expected flux density (calculated as described in \S\ref{sec:methods}) normalized by the value of the effective sensitivity curve for that pulsar, so larger points indicate pulsars with expected flux density much higher than the minimum detectable flux density at the pulsar's position.

In total, there are 116 undetected pulsars plotted in Figure \ref{fig:p0vsdm} that have been classified as ``unexpected" by the logic above. Many of these pulsars are quite close to the sensitivity line, and so small errors in other flux density measurements and spectral indices may change them to ``expected." Because the effective sensitivity curve includes temperature and bandwidth (RFI, by proxy) information, reasons for missed detections are limited to effects that are harder to characterize. The most likely contributors include scintillation, abnormal pulsar behavior (i.e. nulling), and imprecise previous measurements of pulsar parameters resulting in inflated expected flux densities. Scintillation depends on DM \citep{cl91}, with increased timescales for smaller DM. Many of the non-detected pulsars that are outside of the Galactic plane are in this low-DM high-scintillation regime, and are likely to have been obscured (the expected number of scintles in the observation are on the order of $\sim$10). Many of the other missed detections, especially those from surveys with comparable limiting fluxes, were inspected individually. Many of these were obscured by significant RFI across the band. For example, PSR J0108$-$1431 (spin period of $\simeq0.81$\,s and DM of 2.38\,pc\, cm$^{-3}$, to the right of the bottom center of Figure \ref{fig:p0vsdm}) should be easily detected but was obscured by RFI. When examining a number of the other sources, it was found that many of the published spectral indices came from a 1400\,MHz study conducted by \cite{hwx+17}, and were unusually steep. This steepness results in high expected values of flux at 350\,MHz, which are not reflected in our results.

\subsection{RFI Analysis}\label{sec:rfi}
\begin{figure}
\epsscale{1.2}
\plotone{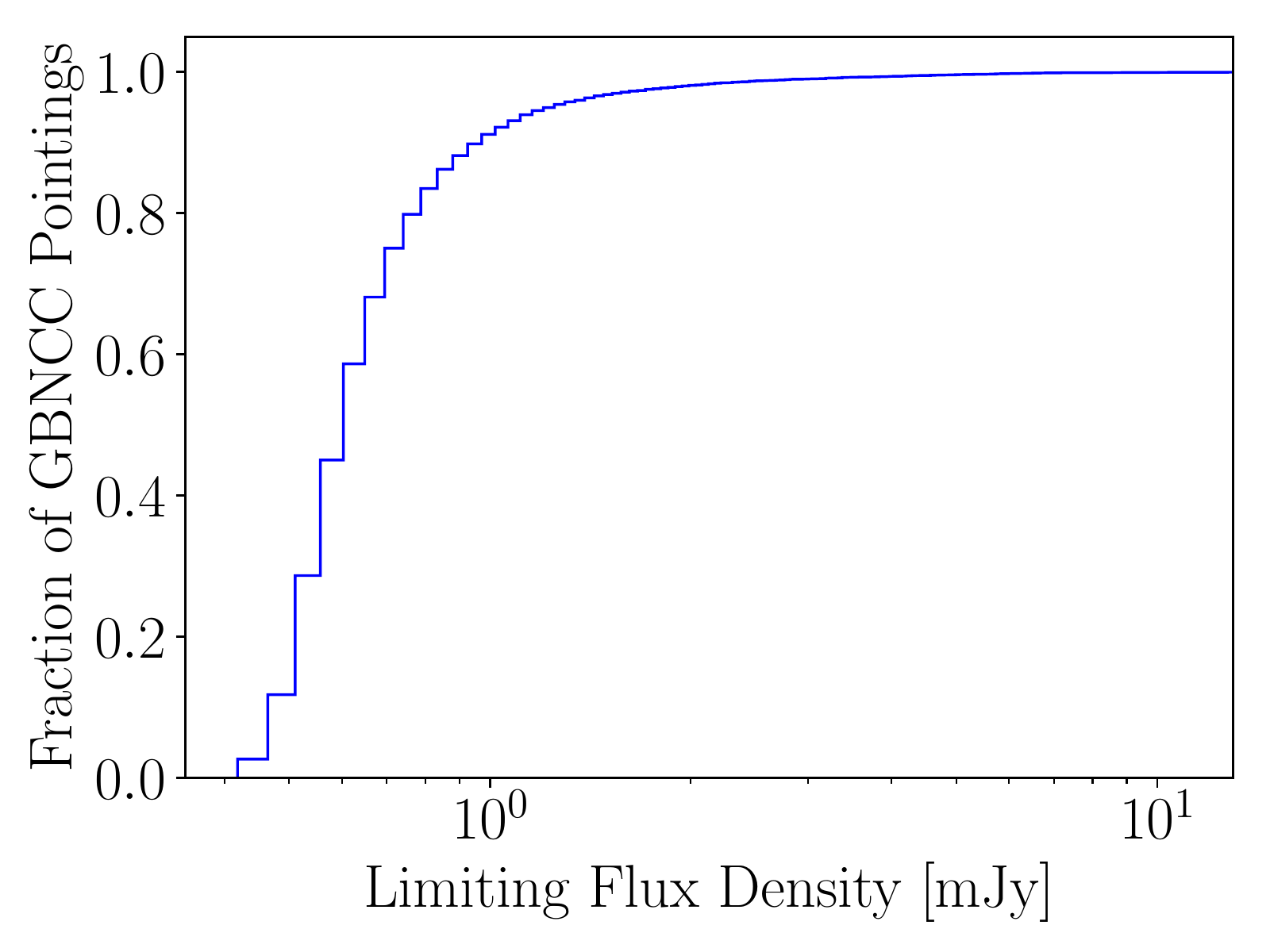}
    \caption{Cumulative histogram of limiting flux density for GBNCC. The mean and median limiting flux densities in the histogram 0.74\,mJy and 0.62\,mJy, and the values range from 0.42\,mJy to 47.\,mJy. All flux density values are given in mJy.}
    \label{fig:flux_lim_hist}
\end{figure}

\begin{figure*}
\plotone{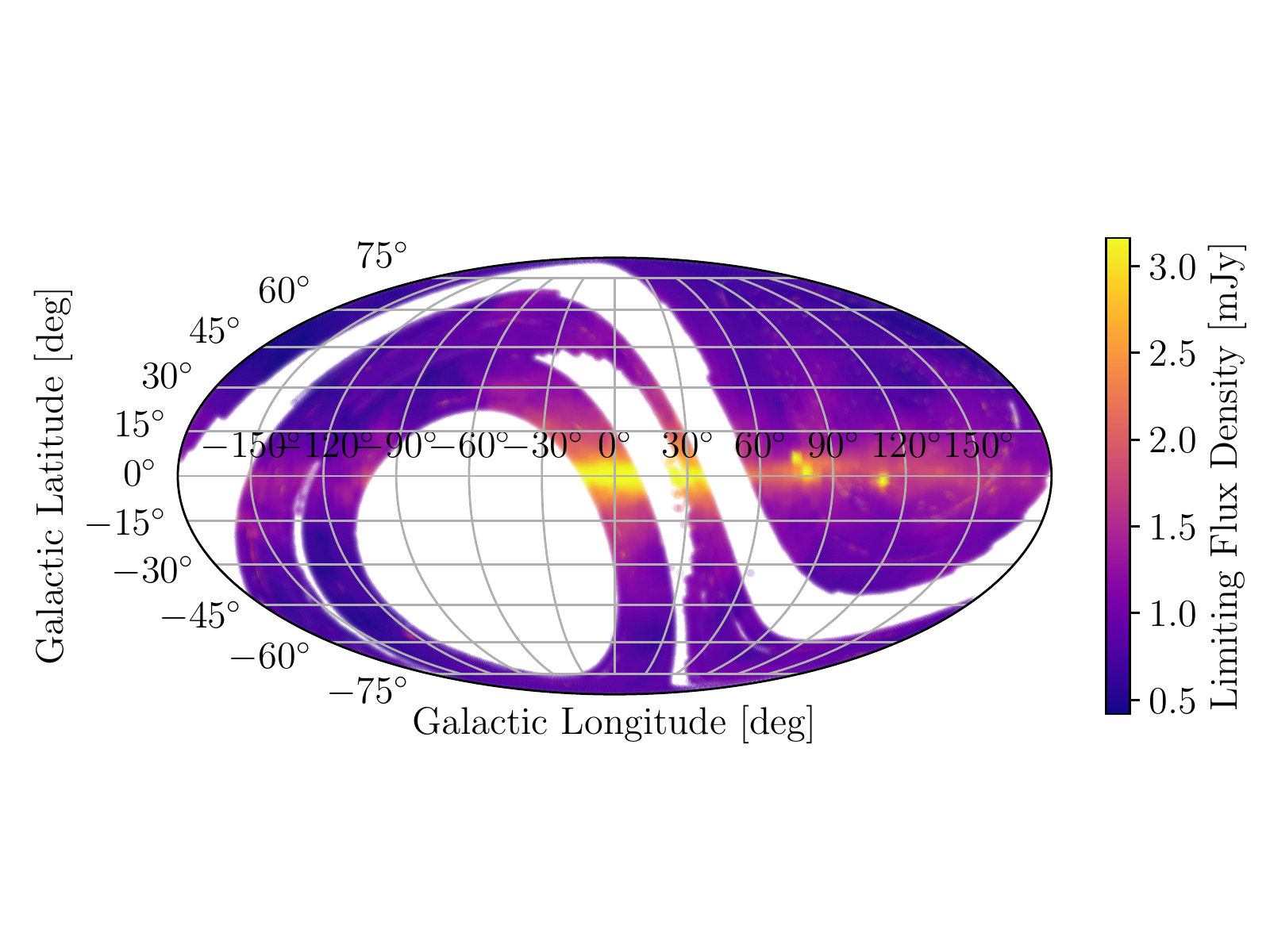}
    \caption{Sky map of GBNCC beams, colored by limiting flux density. The map is plotted in Galactic coordinates on a Mollweide projection, and the flux density is given in mJy.}
    \label{fig:flux_lim_map}
\end{figure*}

To visualize how RFI affects the efficiency of the survey, we determined the limiting flux density for each beam based on a S/N cutoff of 3.8, the temperature at the sky position of the beam, and the bandwidth available after RFI excision. Figure \ref{fig:flux_lim_hist} displays a histogram of the beams by their limiting flux, and Figure \ref{fig:flux_lim_map} shows these same data projected onto their sky positions. The sky map depicts a few important characteristics of the survey: the most obvious is the decreased sensitivity near the Galactic plane, but also visible are many individual pointings within the completed regions where significant RFI masking has reduced sensitivity. To mitigate this, these beams will be scheduled for reobserving. There is a small discrepancy between the number of observed beams displayed in Figures \ref{fig:skymap} and \ref{fig:flux_lim_map} due to a backlog of data which has yet to processed, and so mask fractions have not been determined for these beams.

\subsection{Nulling/Mode-Changing Candidates}
The large set of data analyzed in this study as well as the ``by-eye" verification of all detections allowed for easy identification of potential nulling/mode-changing candidates in the results. This way, we are sensitive to nulling timescales between that of the pulsar spin period and the observation time (120 seconds). These cases were first identified by the appearance of missing pulses in the time-phase plots from processing using the \texttt{PRESTO} package. When a pulsar was noted as a candidate, we followed up using the \texttt{dspsr}\footnote{\url{http://dspsr.sourceforge.net/index.shtml}} package. We folded the time series data in 10 second integrations, zapped remaining RFI by hand, and integrated across frequency using the \texttt{pav} and \texttt{pam} commands within \texttt{PSRCHIVE}\footnote{\url{http://psrchive.sourceforge.net/index.shtml}}. When it was possible to discern on- and off-pulse regions by eye (i.e. significant changes in intensity for some rotations), the candidates were considered likely to be nulling. Some pulsars exhibited behavior similar to mode-changing, where multiple components of the averaged profile were found to be on during different portions of the observation. These pulsars were not treated differently than other nulling candidates $-$ we folded for single pulses to determine the likelihood that different components were visible. All of these sources will be followed up in later works regarding these data. In total, 223 pulsars were found to exhibit intensity variations similar to nulling or mode-changing during their observations, 62 of which have not previously been found to to do so. These candidates' names are marked in Table \ref{tbl:det} with an asterisk.

\subsection{The Galactic Pulsar Population}
Given its overall sky coverage and the large number of pulsar detections reported here (670), the GBNCC survey will play an important role in future understanding of the Galactic pulsar population. To date, the GBNCC survey has detected 571 non-recycled (long-period) pulsars in the Galactic field and 70 Galactic MSPs, which have undergone recycling and have spin periods, $P<30$\,ms. Remaining detections are either associated with globular clusters (3) or are recycled pulsars with spin periods, $P>30$\,ms (26), and have been intentionally ignored for the following analysis, since our current models do not adequately describe the features of this sub-population.

To estimate expected numbers of non-recycled/millisecond pulsar detections in the GBNCC survey, Galactic populations were simulated using {\sc PsrPopPy2}\footnote{https://github.com/devanshkv/PsrPopPy2}, a more recent and currently-maintained version of {\sc PsrPopPy} \citep[][and citations within]{blr+14}. Pulsar populations were generated using {\sc PsrPopPy2}'s {\tt populate} function, which simulates pulsars by drawing parameters from predefined distributions until some condition is met. Due to its large sample size, population estimates from the Parkes Multibeam Pulsar Survey (PMPS) provide the best-known sample parameters. For this reason, these results were used to set a limit on the number of pulsars simulated by {\tt populate}. For the non-recycled pulsar population, pulsars were generated until a synthetic PMPS ``detected" 1038 sources; for MSPs, the desired population size was set to 30,000 sources. Specific parameters defining pulsars' Galactic radial distribution, as well as scale height, spin period, luminosity, and duty cycle can be found in \citet{slm+14}. However, an updated model for the MSP $P$-distribution \citep{lem+15} was implemented in simulations here.

Synthetic surveys were conducted with 100 realizations each of the Galactic non-recycled/millisecond pulsar populations using {\tt survey} and a GBNCC model file, including survey parameters identical to those presented in \S\ref{sec:methods} and lists of completed/remaining GBNCC pointing positions. In the first round of simulations, we fixed the S/N limit for detections to ${\rm S/N}_{\rm cut}=3.8$ (as determined in \S\ref{subsec:efficiency}). This simulation predicted 1442/126 simulated detections for non-recycled/millisecond pulsar populations, respectively (on average; compared to 571/70 actual detections). We then fixed the number of simulated non-recycled/millisecond pulsar detections to their actual values (571/70) and found nominal S/N thresholds for each sub-population, ${\rm S/N}_{\rm cut}=15.3/9.1$. The discrepancies between simulated and actual yields suggest uncertainties in population parameters informed primarily by the PMPS survey, which targeted the Galactic plane and was conducted at 1.4\,GHz. Population parameters determined by these previous surveys produce over-estimates for GBNCC pulsar yields. As an all-sky, low-frequency search, the GBNCC survey (when complete) will be a valuable counterpoint to further refine non-recycled/millisecond pulsar population parameters. As we will show below, positional and rotational parameters of the simulated populations do not match the detected population when these thresholds are set.

\begin{figure*}
\centering
\includegraphics[width=0.75\textwidth]{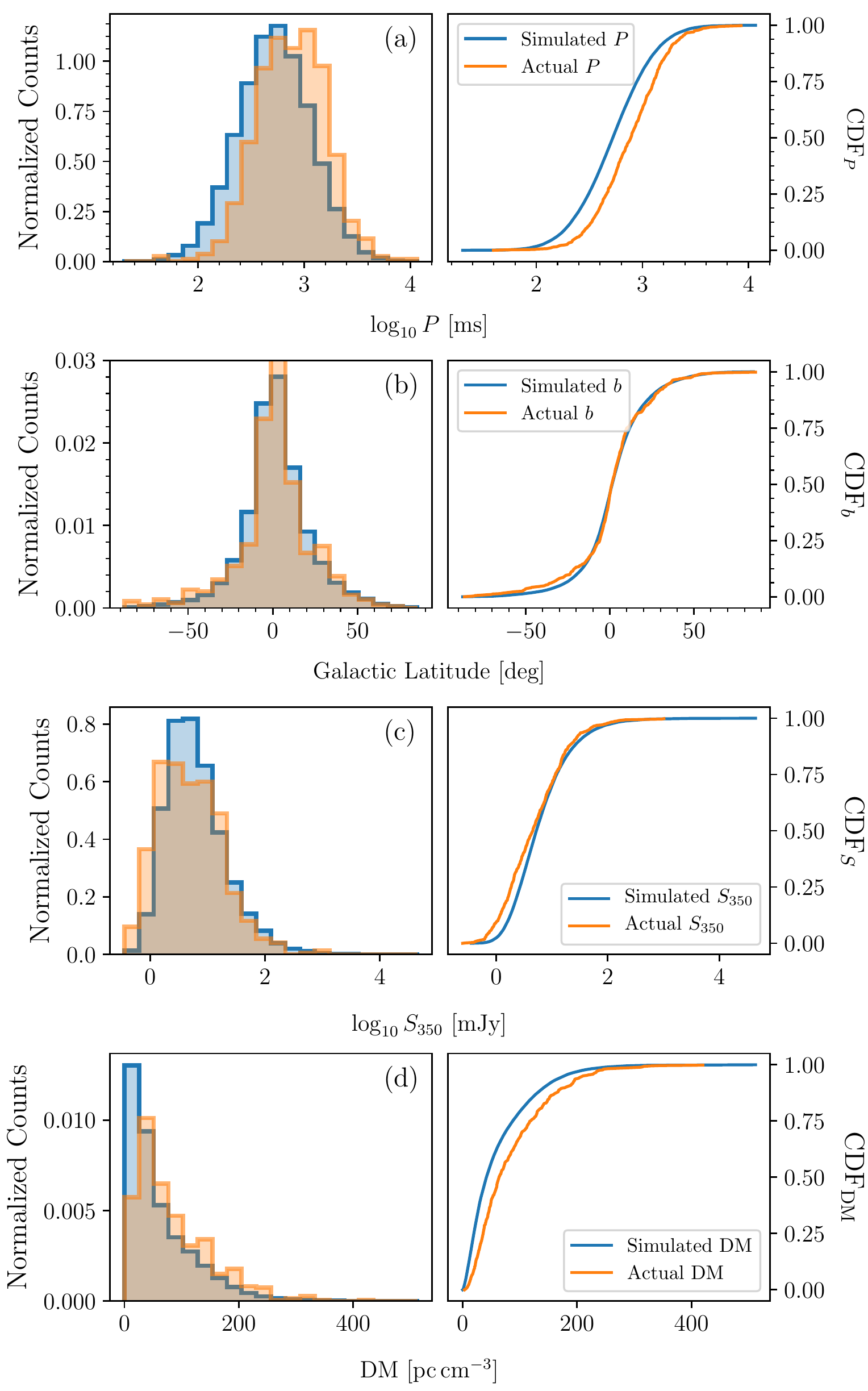}
    \caption{Normalized histograms showing comparisons between (a) spin period, $P$, (b) Galactic latitude, $b$, (c) flux density, $S_{350}$, and (d) dispersion measure, DM, distributions for simulated non-recycled pulsars (blue) and actual detections (orange). The rightmost panel in each row compares actual/simulated CDFs for each parameter. K-S tests comparing these CDFs (see Table \ref{tbl:ks} for details) show disagreement between {\tt act}/{\tt sim} $P$, $S_{350}$, and DM distributions, but $p=41\%$ for $b$ distributions.}
    \label{fig:norm_cdfs}
\end{figure*}
\begin{figure*}
\centering
\includegraphics[width=0.75\textwidth]{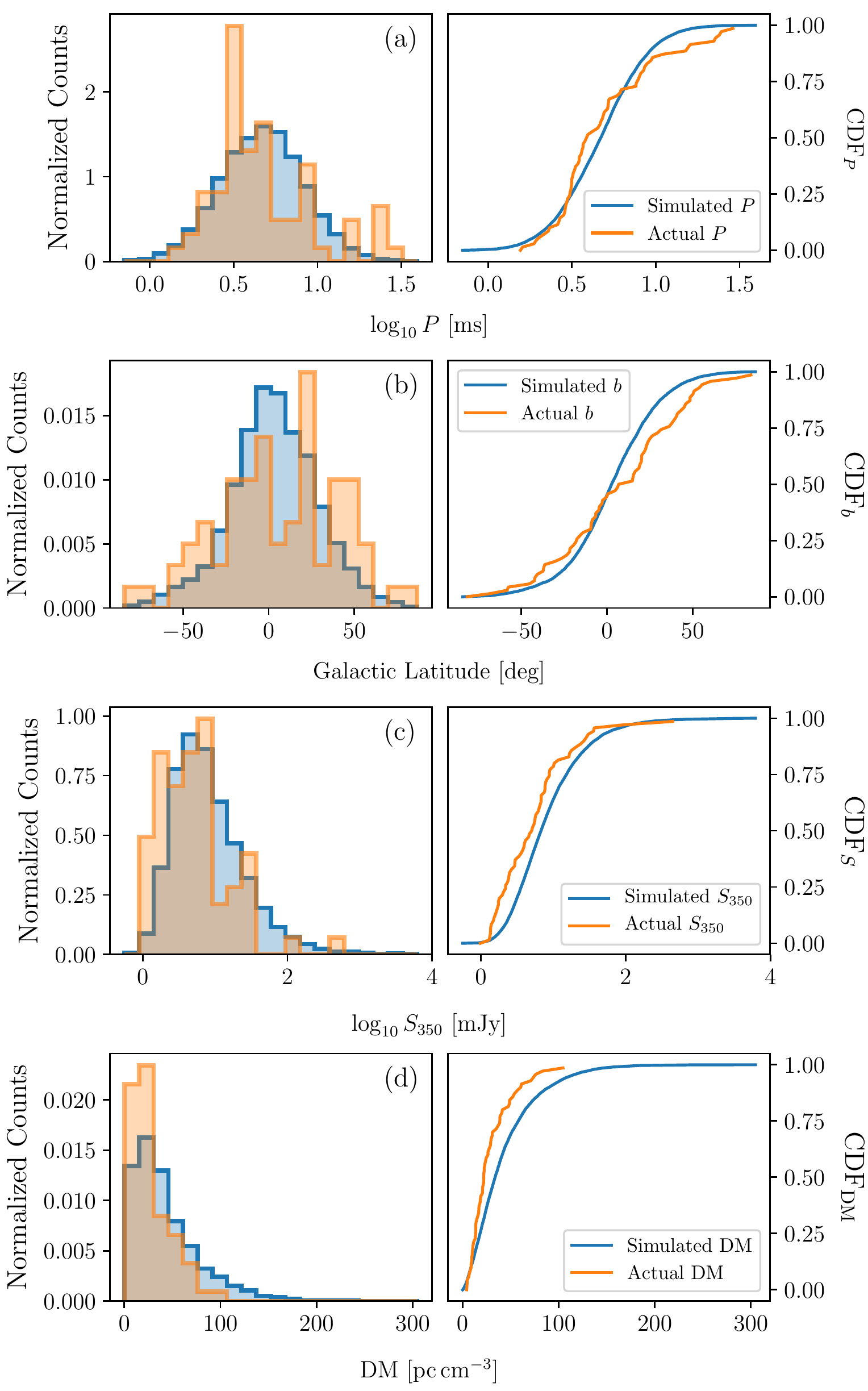}
    \caption{Normalized histograms showing comparisons between (a) spin period, $P$, (b) Galactic latitude, $b$, (c) flux density, $S_{350}$, and (d) dispersion measure, DM, distributions for simulated millisecond pulsars (blue) and actual detections (orange). The rightmost panel in each row compares actual/simulated CDFs for each parameter. K-S tests comparing these CDFs (see Table \ref{tbl:ks} for details) show disagreement between {\tt act}/{\tt sim} $S_{350}$ and DM distributions, but distributions for $b$ and $P$ have $p=3\%$ and $10\%$, respectively.}
    \label{fig:msp_cdfs}
\end{figure*}

\begin{deluxetable}{lcccc}
\tablecaption{K-S test statistics and $p$-values resulting from comparisons between actual/simulated parameter distributions for non-recycled/millisecond pulsars. In cases where the $p$-value is $<1\%$, the null hypothesis (that the two distributions are the same) is rejected.\label{tbl:ks}}
\tablehead{\colhead{Parameter} & \multicolumn{2}{c}{Normal\tablenotemark{a}}& \multicolumn{2}{c}{MSP\tablenotemark{b}}\\
\cline{2-3} \cline{4-5}
& \colhead{K-S} & \colhead{$p(\%)$} & \colhead{K-S} & \colhead{$p(\%)$}
}
\startdata
Spin Period ($P$) & 0.20 & $\ll1$ & 0.14 & \phn$10$ \\ 
Dispersion Measure (DM) & 0.21 & $\ll1$ & 0.26 & $<1$ \\
Flux Density ($S_{350}$) & 0.13 & $\ll1$ & 0.21 & $<1$ \\ 
Galactic Latitude ($b$) & 0.04 & \phn\phn$41$ & 0.17 & \phn\phn$3$ \\
\enddata
\tablenotetext{a}{For simulated non-recycled pulsars, ${\rm S/N}_{\rm cut}=15.3$.}
\tablenotetext{b}{For simulated MSP population, ${\rm S/N}_{\rm cut}=9.1$.}
\end{deluxetable}
To test the validity of underlying non-recycled/millisecond pulsar populations, we compared cumulative distribution functions (CDFs) of simulated ({\tt sim}) pulsar parameters ($P$, DM, $S_{350}$, and $b$) with those of the actual ({\tt act}) detections using a {\tt scipy} implementation of the 2-sample Kolmogorov–Smirnov (K-S) test. For each parameter, the K-S test statistic and $p$-value were computed over a range of ${\rm S/N}_{\rm cut}$. When $p<1\%$, the null hypothesis (that {\tt act}/{\tt sim} parameters are drawn from the same underlying distribution) is rejected. Figures \ref{fig:norm_cdfs} and \ref{fig:msp_cdfs} illustrate these comparisons for non-recycled and millisecond pulsar population parameter distributions, and Table \ref{tbl:ks} summarizes K-S test results when the nominal ${\rm S/N}_{\rm cut}$ values for non-recycled/millisecond pulsar sub-populations (15.3/9.1) are implemented, though we measured these p-values for a range of imposed ${\rm S/N}_{\rm cut}$ values.

Comparing {\tt act}/{\tt sim} parameters for the non-recycled pulsar population, we find broad agreement between $b$ distributions, regardless of ${\rm S/N}_{\rm cut}$. Results for other non-recycled pulsar parameters in Table \ref{tbl:ks} show significant inconsistencies between {\tt act}/{\tt sim} samples. DM distributions are clearly different for ${\rm S/N}_{\rm cut}>4$, likely due to an over-abundance of low-DM simulated detections. For ${\rm S/N}_{\rm cut}=15.3$, we find twice as many {\tt sim} detections with ${\rm DM}<35$\,pc\,cm$^{-3}$. Presumably due to the prevalence of nearby {\tt sim} sources, this sample also has a larger fraction of high-flux density sources, so $S_{350}$ distributions are statistically different for ${\rm S/N}_{\rm cut}=15.3$. However, there is a small window ($10.25<{\rm S/N}_{\rm cut}<12.25$) where {\tt act}/{\tt sim} $S_{350}$ distributions become statistically similar, with $p>1\%$. The null hypothesis is rejected for $P$ due to {\tt act}/{\tt sim} log-normal distributions having different mean values: $\langle\log P_{\rm act}\rangle = 2.88$ versus $\langle\log P_{\rm sim}\rangle = 2.72$ (see Figure \ref{fig:norm_cdfs}). This discrepancy persists, regardless of chosen ${\rm S/N}_{\rm cut}$.

\begin{figure}
\epsscale{1.2}
\plotone{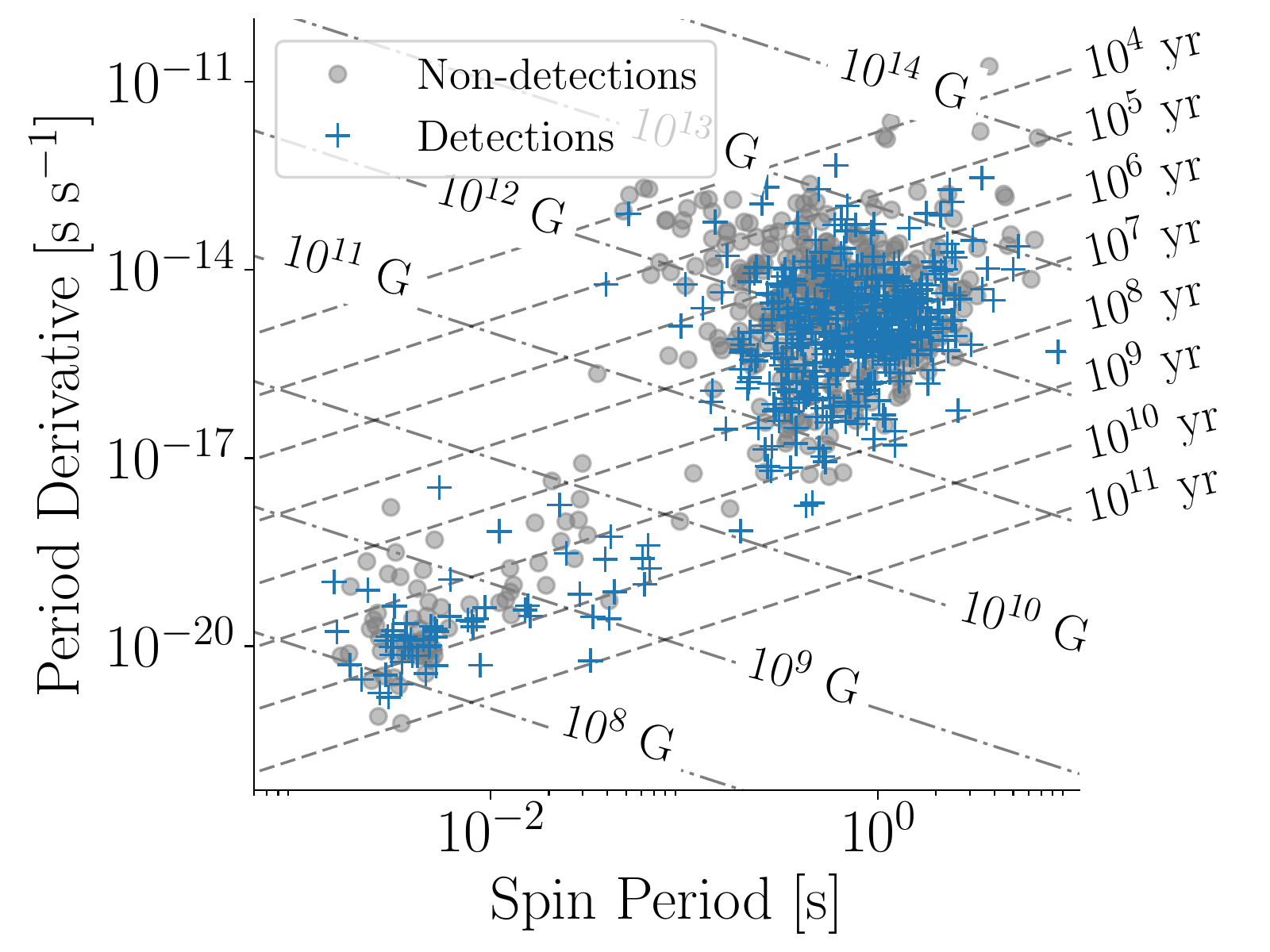}
    \caption{Period vs.\ period derivative for pulsars in GBNCC survey area. Shown in grey are pulsars that were not detected, and blue plus symbols show detections.}
    \label{fig:ppdot}
\end{figure}

Because the simulated versions of the non-recycled pulsar population were primarily informed by PMPS \citep[e.g.][]{lfl+06}, which was conducted at 1.4\,GHz and exclusively covered regions of sky near the Galactic plane ($|b| < 5\,\degr$), we expect there to be bias toward highly dispersed pulsars near the plane. Due to more uniform sky coverage and \--- near the Galactic plane \--- higher sky temperatures and more significant scattering at 350\,MHz, the majority of GBNCC detections ($67\%$) are away from the plane ($|b| > 5\,\degr$). Young pulsars are typically born in the plane and tend to be found nearby, therefore GBNCC's reduced sensitivity to low-latitude sources means that relatively few detections are young pulsars. The $P$-$\dot{P}$ diagram in Figure \ref{fig:ppdot} nicely illustrates this shortage of pulsars detected with characteristic ages, $\tau \leq 1$\,Myr. By imposing an age cutoff on non-recycled pulsars in the ATNF catalog, $\tau > 1$\,Myr, the resulting simulated spin period distribution is statistically similar to that of GBNCC detections (K-S $p>1\%$). This selection effect accounts for the apparent differences between {\tt act}/{\tt sim} $P$-distributions, but can not explain discrepancies in $S_{350}$ and DM distributions for non-recycled pulsars.

K-S tests comparing {\tt act}/{\tt sim} parameter distributions for the MSP population show better agreement (see Table \ref{tbl:ks} and Figure \ref{fig:msp_cdfs}). For MSPs, selection effects based on Galactic latitude and spin period do not come into play since MSPs are more isotropically distributed and model parameters for this sub-population are based on results from multiple Parkes Telescope surveys (see \cite{lem+15}, and references therein). For these reasons, the simulated population's spin periods are statistically similar to the sample detected by GBNCC. This conclusion does not change, regardless of the chosen ${\rm S/N}_{\rm cut}$ value. For $b$, the null hypothesis is still not rejected by our criteria ($p<1\%$). Based on the $b$ histograms themselves, there appears to be an absence of detections in the {\tt act} sample in/near the Galactic plane, which is not the case for {\tt sim} sources. The null hypothesis is rejected for $S_{350}$ due to the over-abundance of high-flux-density sources in the {\tt sim} sample compared to those present in the {\tt act} sample. Median flux densities for {\tt act}/{\tt sim} detections are 4.9/6.8\,mJy respectively. Comparing {\tt act}/{\tt sim} DM distributions, the {\tt sim} sample consists of a higher fraction of high-DM MSPs and 12\% of simulated detections have DMs in excess of the {\tt act} maximum value, $104.5$\,pc\,cm$^{-3}$. This is likely related to the bias toward high-flux-density detections noted in $S_{350}$ for {\tt sim} MSPs mentioned above.

Based on discrepancies between predicted yields from simulations and actual numbers of detections by the GBNCC survey, it appears that model parameters need to be further refined in order to generate more realistic Galactic pulsar populations in the future. For now, we proceed with nominal ${\rm S/N}_{\rm cut}$ values in order to estimate the GBNCC survey's future yields. In the remaining $\approx$21,000 pointings, we expect an additional 160/16 non-recycled/millisecond pulsar detections, or $\approx60/5$ discoveries, accounting for detectable known pulsars in regions of sky remaining \citep{mhth05}.

\section{Conclusions}
\label{sec:conc}
We have provided all detections of currently known pulsars that exist within the area of the 350\,MHz GBNCC pulsar survey and performed some preliminary analysis of the resulting data set. Specifically, we have provided new flux density and pulse width measurements as well as pulse profiles for the 670 detections. When possible, we used our flux density measurements with previous measurements at different frequencies to refine spectral index. We also made a measurement of the spin period-pulse width relation, observing a powerlaw correlation of the form $W_{10} \propto P^{-0.27}$. The low frequency of the survey provides increased sensitivity to dispersion, allowing for more precise measurements of DM for many pulsars that have only been measured in high frequency surveys. Using all of this information, we have made quantitative measurements of the survey's efficacy and the RFI environment at 350\,MHz, with a minimum detectable S/N of $\sim$3.8 and a mean limiting flux density of 0.74\,mJy. These measurements have allowed us to make realistic predictions about the survey's yield when complete based on the detectability of known pulsars in the dataset, and we expect to detect on the order of 160 non-recycled pulsars and 15 MSPs. The simulations from which these expectations come uncovered discrepancies in DM, spin period, and spatial distribution in the Galaxy for the simulated populations which will be addressed in a future study. Combing through the data following processing has brought many interesting characteristics of pulsars in the survey to light, including 223 pulsars exhibiting evidence of variable intensities suggestive of nulling/mode-changing and 4 showing evidence for broken power-law spectral energy distributions. These kinds of qualitative observations pave the way for follow-up quantitative analyses of these data and the remaining beams that will be observed in the next few years. 

\section*{Acknowledgements}
We thank our anonymous referee for their suggestions and guidance. This work was supported by the NANOGrav Physics Frontiers Center, which is supported by the National Science Foundation award 1430284. The Green Bank Observatory is a facility of the National Science Foundation (NSF) operated under cooperative agreement by Associated Universities, Inc. 
R.S.\ acknowledges support through the Australian Research Council grant FL150100148.
WF acknowledges the WVU STEM Mountains of Excellence Graduate Fellowship. MM and MS acknowledge the National Science Foundation OIA award number 1458952. JvL acknowledges funding from the European Research Council under the European Union’s Seventh Framework Programme (FP/2007-2013) / ERC Grant Agreement n. 617199 (`ALERT'), and from Vici research programme `ARGO' with project number 639.043.815, financed by the Netherlands Organisation for Scientific Research (NWO). VMK acknowledges the NSERC Discovery Grant, the Herzberg Award, FRQNT and CRAQ, Canada Research Chairs, CIFAR and the Webster Foundation Fellowship, and the Trottier Chair in Astrophysics and Cosmology. Computations were made on the supercomputer Guillimin at McGill University\footnote{\url{www.hpc.mcgill.ca}}, managed by Calcul Quebec and Compute Canada. The operation of this supercomputer is funded by the Canada Foundation for Innovation (CFI), NanoQuebec, RMGA and the Fonds de recherche du Quebec - Nature et technologies (FRQ-NT). The CyberSKA project was funded by a CANARIE NEP-2 grant. PC acknowledges the FRQNT Doctoral Research Award. SMR is a CIFAR Senior Fellow. Pulsar research at UBC is supported by an NSERC Discovery Grant and by the Canada Foundation for Innovation.

\software{\texttt{Astropy} \citep{astropy:2018}, \texttt{PRESTO} \citep{smr+01},  
\texttt{PsrPopPy2} \citep{blr+14}, 
\texttt{SciPy} \citep{scipy}, 
\texttt{NumPy} \citep{numpy}, 
\texttt{dspsr} \citep{sb11}, 
\texttt{PSRCHIVE} \citep{hsm04}, 
\texttt{TEMPO} (\url{http://tempo.sourceforge.net/})}

\facilities{Robert C. Byrd Green Bank Telescope (GBT)}

\bibliographystyle{apj}
\bibliography{gbncc}

\begin{thebibliography}{107}
\expandafter\ifx\csname natexlab\endcsname\relax\def\natexlab#1{#1}\fi

\bibitem[{bbf(1984)}]{bbf+84}
 1984, {Birth and evolution of neutron stars: Issues raised by millisecond
  pulsars}

\bibitem[{{Aloisi} {et~al.}(2019){Aloisi}, {Cruz}, {Daniels}, {Meyers},
  {Roekle}, {Schuett}, {Swiggum}, {DeCesar}, {Kaplan}, {Lynch}, {Stovall},
  {Levin}, {Archibald}, {Banaszak}, {Biwer}, {Boyles}, {Chawla}, {Dartez},
  {Cui}, {Day}, {Ford}, {Flanigan}, {Fonseca}, {Hessels}, {Hinojosa},
  {Karako-Argaman}, {Kaspi}, {Kondratiev}, {Leake}, {Lunsford}, {Martinez},
  {Mata}, {McLaughlin}, {Noori}, {Ransom}, {Roberts}, {Rohr}, {Siemens},
  {Spiewak}, {Stairs}, {van Leeuwen}, {Walker}, \& {Wells}}]{acd+19}
{Aloisi}, R.~J., {et~al.} 2019, \apj, in press, arXiv:1903.03543

\bibitem[{{Ashworth} \& {Lyne}(1981)}]{al81}
{Ashworth}, M., \& {Lyne}, A.~G. 1981, \mnras, 195, 517

\bibitem[{{Bailes} {et~al.}(1997){Bailes}, {Johnston}, {Bell}, {Lorimer},
  {Stappers}, {Manchester}, {Lyne}, {Nicastro}, \& {Gaensler}}]{bjb+97}
{Bailes}, M., {et~al.} 1997, \apj, 481, 386

\bibitem[{{Barr} {et~al.}(2013){Barr}, {Champion}, {Kramer}, {Eatough},
  {Freire}, {Karuppusamy}, {Lee}, {Verbiest}, {Bassa}, {Lyne}, {Stappers},
  {Lorimer}, \& {Klein}}]{bck+13}
{Barr}, E.~D., {et~al.} 2013, \mnras, 435, 2234

\bibitem[{{Bates} {et~al.}(2014){Bates}, {Lorimer}, {Rane}, \&
  {Swiggum}}]{blr+14}
{Bates}, S.~D., {Lorimer}, D.~R., {Rane}, A., \& {Swiggum}, J. 2014, \mnras,
  439, 2893

\bibitem[{{Bell} {et~al.}(2016){Bell}, {Murphy}, {Johnston}, {Kaplan}, {Croft},
  {Hancock}, {Callingham}, {Zic}, {Dobie}, {Swiggum}, {Rowlinson},
  {Hurley-Walker}, {Offringa}, {Bernardi}, {Bowman}, {Briggs}, {Cappallo},
  {Deshpand e}, {Gaensler}, {Greenhill}, {Hazelton}, {Johnston-Hollitt},
  {Lonsdale}, {McWhirter}, {Mitchell}, {Morales}, {Morgan}, {Oberoi}, {Ord},
  {Prabu}, {Shankar}, {Srivani}, {Subrahmanyan}, {Tingay}, {Wayth}, {Webster},
  {Williams}, \& {Williams}}]{bmj+16}
{Bell}, M.~E., {et~al.} 2016, \mnras, 461, 908

\bibitem[{{Bhattacharyya} {et~al.}(2013){Bhattacharyya}, {Roy}, {Ray}, {Gupta},
  {Bhattacharya}, {Romani}, {Ransom}, {Ferrara}, {Wolff}, {Camilo}, {Cognard},
  {Harding}, {den Hartog}, {Johnston}, {Keith}, {Kerr}, {Michelson}, {Saz
  Parkinson}, {Wood}, \& {Wood}}]{brr+13}
{Bhattacharyya}, B., {et~al.} 2013, \apjl, 773, L12

\bibitem[{{Biggs} {et~al.}(1994){Biggs}, {Bailes}, {Lyne}, {Goss}, \&
  {Fruchter}}]{bbl+94}
{Biggs}, J.~D., {Bailes}, M., {Lyne}, A.~G., {Goss}, W.~M., \& {Fruchter},
  A.~S. 1994, \mnras, 267, 125

\bibitem[{{Bilous} {et~al.}(2016){Bilous}, {Kondratiev}, {Kramer}, {Keane},
  {Hessels}, {Stappers}, {Malofeev}, {Sobey}, {Breton}, {Cooper}, {Falcke},
  {Karastergiou}, {Michilli}, {Os{\l}owski}, {Sanidas}, {ter Veen}, {van
  Leeuwen}, {Verbiest}, {Weltevrede}, {Zarka}, {Grie{\ss}meier}, {Serylak},
  {Bell}, {Broderick}, {Eisl{\"o}ffel}, {Markoff}, \& {Rowlinson}}]{bkk+16}
{Bilous}, A.~V., {et~al.} 2016, \aap, 591, A134

\bibitem[{{Boyles} {et~al.}(2013){Boyles}, {Lynch}, {Ransom}, {Stairs},
  {Lorimer}, {McLaughlin}, {Hessels}, {Kaspi}, {Kondratiev}, {Archibald},
  {Berndsen}, {Cardoso}, {Cherry}, {Epstein}, {Karako-Argaman}, {McPhee},
  {Pennucci}, {Roberts}, {Stovall}, \& {van Leeuwen}}]{blr+13}
{Boyles}, J., {et~al.} 2013, \apj, 763, 80

\bibitem[{{Brinkman} {et~al.}(2018){Brinkman}, {Freire}, {Rankin}, \&
  {Stovall}}]{bfrs18}
{Brinkman}, C., {Freire}, P. C.~C., {Rankin}, J., \& {Stovall}, K. 2018,
  \mnras, 474, 2012

\bibitem[{{Burgay} {et~al.}(2006){Burgay}, {Joshi}, {D'Amico}, {Possenti},
  {Lyne}, {Manchester}, {McLaughlin}, {Kramer}, {Camilo}, \& {Freire}}]{bjd+06}
{Burgay}, M., {et~al.} 2006, \mnras, 368, 283

\bibitem[{{Burgay} {et~al.}(2013){Burgay}, {Keith}, {Lorimer}, {Hassall},
  {Lyne}, {Camilo}, {D'Amico}, {Hobbs}, {Kramer}, {Manchester}, {McLaughlin},
  {Possenti}, {Stairs}, \& {Stappers}}]{bkl+13}
---. 2013, \mnras, 429, 579

\bibitem[{{Camilo}(1995)}]{cam95a}
{Camilo}, F. 1995, PhD thesis, PRINCETON UNIVERSITY.

\bibitem[{{Camilo} \& {Nice}(1995)}]{cn95}
{Camilo}, F., \& {Nice}, D.~J. 1995, \apj, 445, 756

\bibitem[{{Camilo} {et~al.}(1996{\natexlab{a}}){Camilo}, {Nice}, {Shrauner}, \&
  {Taylor}}]{cnst96}
{Camilo}, F., {Nice}, D.~J., {Shrauner}, J.~A., \& {Taylor}, J.~H.
  1996{\natexlab{a}}, \apj, 469, 819

\bibitem[{{Camilo} {et~al.}(1996{\natexlab{b}}){Camilo}, {Nice}, \&
  {Taylor}}]{cnt96}
{Camilo}, F., {Nice}, D.~J., \& {Taylor}, J.~H. 1996{\natexlab{b}}, \apj, 461,
  812

\bibitem[{{Champion} {et~al.}(2005){Champion}, {Lorimer}, {McLaughlin},
  {Xilouris}, {Arzoumanian}, {Freire}, {Lommen}, {Cordes}, \&
  {Camilo}}]{clm+05}
{Champion}, D.~J., {et~al.} 2005, \mnras, 363, 929

\bibitem[{{Chandler}(2003)}]{cha03}
{Chandler}, A.~M. 2003, PhD thesis, CALIFORNIA INSTITUTE OF TECHNOLOGY

\bibitem[{{Chatterjee} {et~al.}(2005){Chatterjee}, {Goss}, \&
  {Brisken}}]{cgb05}
{Chatterjee}, S., {Goss}, W.~M., \& {Brisken}, W.~F. 2005, \apjl, 634, L101

\bibitem[{{Chen} \& {Wang}(2014)}]{cw14}
{Chen}, J.~L., \& {Wang}, H.~G. 2014, \apjs, 215, 11

\bibitem[{{Coenen} {et~al.}(2014){Coenen}, {van Leeuwen}, {Hessels},
  {Stappers}, {Kondratiev}, {Alexov}, {Breton}, {Bilous}, {Cooper}, {Falcke},
  {Fallows}, {Gajjar}, {Grie{\ss}meier}, {Hassall}, {Karastergiou}, {Keane},
  {Kramer}, {Kuniyoshi}, {Noutsos}, {Os{\l}owski}, {Pilia}, {Serylak},
  {Schrijvers}, {Sobey}, {ter Veen}, {Verbiest}, {Weltevrede}, {Wijnholds},
  {Zagkouris}, {van Amesfoort}, {Anderson}, {Asgekar}, {Avruch}, {Bell},
  {Bentum}, {Bernardi}, {Best}, {Bonafede}, {Breitling}, {Broderick},
  {Br{\"u}ggen}, {Butcher}, {Ciardi}, {Corstanje}, {Deller}, {Duscha},
  {Eisl{\"o}ffel}, {Fender}, {Ferrari}, {Frieswijk}, {Garrett}, {de Gasperin},
  {de Geus}, {Gunst}, {Hamaker}, {Heald}, {Hoeft}, {van der Horst}, {Juette},
  {Kuper}, {Law}, {Mann}, {McFadden}, {McKay-Bukowski}, {McKean}, {Munk},
  {Orru}, {Paas}, {Pand ey-Pommier}, {Polatidis}, {Reich}, {Renting},
  {R{\"o}ttgering}, {Rowlinson}, {Scaife}, {Schwarz}, {Sluman}, {Smirnov},
  {Swinbank}, {Tagger}, {Tang}, {Tasse}, {Thoudam}, {Toribio}, {Vermeulen},
  {Vocks}, {van Weeren}, {Wucknitz}, {Zarka}, \& {Zensus}}]{cvh+14}
{Coenen}, T., {et~al.} 2014, \aap, 570, A60

\bibitem[{{Cognard} {et~al.}(2011){Cognard}, {Guillemot}, {Johnson}, {Smith},
  {Venter}, {Harding}, {Wolff}, {Cheung}, {Donato}, {Abdo}, {Ballet}, {Camilo},
  {Desvignes}, {Dumora}, {Ferrara}, {Freire}, {Grove}, {Johnston}, {Keith},
  {Kramer}, {Lyne}, {Michelson}, {Parent}, {Ransom}, {Ray}, {Romani}, {Saz
  Parkinson}, {Stappers}, {Theureau}, {Thompson}, {Weltevrede}, \&
  {Wood}}]{cgj+11}
{Cognard}, I., {et~al.} 2011, \apj, 732, 47

\bibitem[{{Cordes} \& {Lazio}(1991)}]{cl91}
{Cordes}, J.~M., \& {Lazio}, T.~J. 1991, \apj, 376, 123

\bibitem[{{Cordes} {et~al.}(2006){Cordes}, {Freire}, {Lorimer}, {Camilo},
  {Champion}, {Nice}, {Ramachandran}, {Hessels}, {Vlemmings}, {van Leeuwen},
  {Ransom}, {Bhat}, {Arzoumanian}, {McLaughlin}, {Kaspi}, {Kasian}, {Deneva},
  {Reid}, {Chatterjee}, {Han}, {Backer}, {Stairs}, {Deshpande}, \&
  {Faucher-Gigu{\`e}re}}]{cfl+06}
{Cordes}, J.~M., {et~al.} 2006, \apj, 637, 446

\bibitem[{{Costa} {et~al.}(1991){Costa}, {McCulloch}, \& {Hamilton}}]{cmh91}
{Costa}, M.~E., {McCulloch}, P.~M., \& {Hamilton}, P.~A. 1991, \mnras, 252, 13

\bibitem[{{Dembska} {et~al.}(2014){Dembska}, {Kijak}, {Jessner}, {Lewandowski},
  {Bhattacharyya}, \& {Gupta}}]{dkj+14}
{Dembska}, M., {Kijak}, J., {Jessner}, A., {Lewandowski}, W., {Bhattacharyya},
  B., \& {Gupta}, Y. 2014, \mnras, 445, 3105

\bibitem[{{Deneva} {et~al.}(2013){Deneva}, {Stovall}, {McLaughlin}, {Bates},
  {Freire}, {Martinez}, {Jenet}, \& {Bagchi}}]{dsm+13}
{Deneva}, J.~S., {Stovall}, K., {McLaughlin}, M.~A., {Bates}, S.~D., {Freire},
  P.~C.~C., {Martinez}, J.~G., {Jenet}, F., \& {Bagchi}, M. 2013, \apj, 775, 51

\bibitem[{{Dewey} {et~al.}(1985{\natexlab{a}}){Dewey}, {Taylor}, {Weisberg}, \&
  {Stokes}}]{dtw+85}
{Dewey}, R.~J., {Taylor}, J.~H., {Weisberg}, J.~M., \& {Stokes}, G.~H.
  1985{\natexlab{a}}, \apjl, 294, L25

\bibitem[{{Dewey} {et~al.}(1985{\natexlab{b}}){Dewey}, {Taylor}, {Weisberg}, \&
  {Stokes}}]{dtws85}
---. 1985{\natexlab{b}}, \apjl, 294, L25

\bibitem[{{Fomalont} {et~al.}(1992){Fomalont}, {Goss}, {Lyne}, {Manchester}, \&
  {Justtanont}}]{fgl+92}
{Fomalont}, E.~B., {Goss}, W.~M., {Lyne}, A.~G., {Manchester}, R.~N., \&
  {Justtanont}, K. 1992, \mnras, 258, 497

\bibitem[{{Foster} {et~al.}(1991){Foster}, {Fairhead}, \& {Backer}}]{ffb91}
{Foster}, R.~S., {Fairhead}, L., \& {Backer}, D.~C. 1991, \apj, 378, 687

\bibitem[{{Frail} {et~al.}(2016){Frail}, {Jagannathan}, {Mooley}, \&
  {Intema}}]{fjmi16}
{Frail}, D.~A., {Jagannathan}, P., {Mooley}, K.~P., \& {Intema}, H.~T. 2016,
  \apj, 829, 119

\bibitem[{{Fruchter} {et~al.}(1990){Fruchter}, {Berman}, {Bower}, {Convery},
  {Goss}, {Hankins}, {Klein}, {Nice}, {Ryba}, {Stinebring}, {Taylor},
  {Thorsett}, \& {Weisberg}}]{fbb+90}
{Fruchter}, A.~S., {et~al.} 1990, \apj, 351, 642

\bibitem[{{Gomez-Gonzalez} \& {Guelin}(1974)}]{gg74}
{Gomez-Gonzalez}, J., \& {Guelin}, M. 1974, \aap, 32, 441

\bibitem[{{Gould} \& {Lyne}(1998)}]{gl98}
{Gould}, D.~M., \& {Lyne}, A.~G. 1998, \mnras, 301, 235

\bibitem[{{Halpern} {et~al.}(2001){Halpern}, {Camilo}, {Gotthelf}, {Helfand },
  {Kramer}, {Lyne}, {Leighly}, \& {Eracleous}}]{hcg+01}
{Halpern}, J.~P., {Camilo}, F., {Gotthelf}, E.~V., {Helfand }, D.~J., {Kramer},
  M., {Lyne}, A.~G., {Leighly}, K.~M., \& {Eracleous}, M. 2001, \apjl, 552,
  L125

\bibitem[{{Han} {et~al.}(2017){Han}, {Wang}, {Xu}, \& {Han}}]{hwx+17}
{Han}, J., {Wang}, C., {Xu}, J., \& {Han}, J. 2017, arXiv e-prints,
  arXiv:1703.05988

\bibitem[{{Han} {et~al.}(2016){Han}, {Wang}, {Xu}, \& {Han}}]{hwxh16}
{Han}, J., {Wang}, C., {Xu}, J., \& {Han}, J.-L. 2016, Research in Astronomy
  and Astrophysics, 16, 159

\bibitem[{{Haslam} {et~al.}(1981){Haslam}, {Klein}, {Salter}, {Stoffel},
  {Wilson}, {Cleary}, {Cooke}, \& {Thomasson}}]{hks+81}
{Haslam}, C.~G.~T., {Klein}, U., {Salter}, C.~J., {Stoffel}, H., {Wilson},
  W.~E., {Cleary}, M.~N., {Cooke}, D.~J., \& {Thomasson}, P. 1981, \aap, 100,
  209

\bibitem[{{Hobbs} {et~al.}(2004){Hobbs}, {Faulkner}, {Stairs}, {Camilo},
  {Manchester}, {Lyne}, {Kramer}, {D'Amico}, {Kaspi}, {Possenti}, {McLaughlin},
  {Lorimer}, {Burgay}, {Joshi}, \& {Crawford}}]{hfs+04}
{Hobbs}, G., {et~al.} 2004, \mnras, 352, 1439

\bibitem[{{Hotan} {et~al.}(2004){Hotan}, {van Straten}, \&
  {Manchester}}]{hsm04}
{Hotan}, A.~W., {van Straten}, W., \& {Manchester}, R.~N. 2004, \pasa, 21, 302

\bibitem[{{Hulse} \& {Taylor}(1975)}]{ht75b}
{Hulse}, R.~A., \& {Taylor}, J.~H. 1975, \apjl, 201, L55

\bibitem[{{Jacoby} {et~al.}(2007){Jacoby}, {Bailes}, {Ord}, {Knight}, \&
  {Hotan}}]{jbo+07}
{Jacoby}, B.~A., {Bailes}, M., {Ord}, S.~M., {Knight}, H.~S., \& {Hotan}, A.~W.
  2007, \apj, 656, 408

\bibitem[{{Jankowski} {et~al.}(2018{\natexlab{a}}){Jankowski}, {van Straten},
  {Keane}, {Bailes}, {Barr}, {Johnston}, \& {Kerr}}]{jsk+18}
{Jankowski}, F., {van Straten}, W., {Keane}, E.~F., {Bailes}, M., {Barr},
  E.~D., {Johnston}, S., \& {Kerr}, M. 2018{\natexlab{a}}, \mnras, 473, 4436

\bibitem[{{Jankowski} {et~al.}(2018{\natexlab{b}}){Jankowski}, {van Straten},
  {Keane}, {Bailes}, {Barr}, {Johnston}, \& {Kerr}}]{jvk+18}
---. 2018{\natexlab{b}}, \mnras, 473, 4436

\bibitem[{{Jankowski} {et~al.}(2019){Jankowski}, {Bailes}, {van Straten},
  {Keane}, {Flynn}, {Barr}, {Bateman}, {Bhandari}, {Caleb}, {Campbell-Wilson},
  {Farah}, {Green}, {Hunstead}, {Jameson}, {Os{\l}owski}, {Parthasarathy},
  {Rosado}, \& {Venkatraman Krishnan}}]{jbv+19}
{Jankowski}, F., {et~al.} 2019, \mnras, 484, 3691

\bibitem[{{Janssen} {et~al.}(2010){Janssen}, {Stappers}, {Bassa}, {Cognard},
  {Kramer}, \& {Theureau}}]{jsb+10}
{Janssen}, G.~H., {Stappers}, B.~W., {Bassa}, C.~G., {Cognard}, I., {Kramer},
  M., \& {Theureau}, G. 2010, \aap, 514, A74

\bibitem[{{Janssen} {et~al.}(2009){Janssen}, {Stappers}, {Braun}, {van
  Straten}, {Edwards}, {Rubio-Herrera}, {van Leeuwen}, \&
  {Weltevrede}}]{jsb+09}
{Janssen}, G.~H., {Stappers}, B.~W., {Braun}, R., {van Straten}, W., {Edwards},
  R.~T., {Rubio-Herrera}, E., {van Leeuwen}, J., \& {Weltevrede}, P. 2009,
  \aap, 498, 223

\bibitem[{{Johnston} \& {Karastergiou}(2019)}]{jk19}
{Johnston}, S., \& {Karastergiou}, A. 2019, \mnras, 485, 640

\bibitem[{{Johnston} \& {Kerr}(2018)}]{jk18}
{Johnston}, S., \& {Kerr}, M. 2018, \mnras, 474, 4629

\bibitem[{{Johnston} {et~al.}(1992){Johnston}, {Lyne}, {Manchester}, {Kniffen},
  {D'Amico}, {Lim}, \& {Ashworth}}]{jlm+92}
{Johnston}, S., {Lyne}, A.~G., {Manchester}, R.~N., {Kniffen}, D.~A.,
  {D'Amico}, N., {Lim}, J., \& {Ashworth}, M. 1992, \mnras, 255, 401

\bibitem[{Jones {et~al.}(2001)Jones, Oliphant, Peterson, {et~al.}}]{scipy}
Jones, E., Oliphant, T., Peterson, P., {et~al.} 2001, {SciPy}: Open source
  scientific tools for {Python}, [Online; accessed <today>]

\bibitem[{{Joshi} {et~al.}(2009){Joshi}, {McLaughlin}, {Lyne}, {Ludovici},
  {Pawar}, {Faulkner}, {Lorimer}, {Kramer}, \& {Davies}}]{jml+09}
{Joshi}, B.~C., {et~al.} 2009, \mnras, 398, 943

\bibitem[{{Karako-Argaman} {et~al.}(2015){Karako-Argaman}, {Kaspi}, {Lynch},
  {Hessels}, {Kondratiev}, {McLaughlin}, {Ransom}, {Archibald}, {Boyles},
  {Jenet}, {Kaplan}, {Levin}, {Lorimer}, {Madsen}, {Roberts}, {Siemens},
  {Stairs}, {Stovall}, {Swiggum}, \& {van Leeuwen}}]{kkl+15}
{Karako-Argaman}, C., {et~al.} 2015, \apj, 809, 67

\bibitem[{{Kawash} {et~al.}(2018){Kawash}, {McLaughlin}, {Kaplan}, {DeCesar},
  {Levin}, {Lorimer}, {Lynch}, {Stovall}, {Swiggum}, {Fonseca}, {Archibald},
  {Banaszak}, {Biwer}, {Boyles}, {Cui}, {Dartez}, {Day}, {Ernst}, {Ford},
  {Flanigan}, {Heatherly}, {Hessels}, {Hinojosa}, {Jenet}, {Karako-Argaman},
  {Kaspi}, {Kondratiev}, {Leake}, {Lunsford}, {Martinez}, {Mata}, {Matheny},
  {Mcewen}, {Mingyar}, {Orsini}, {Ransom}, {Roberts}, {Rohr}, {Siemens},
  {Spiewak}, {Stairs}, {van Leeuwen}, {Walker}, \& {Wells}}]{kmlk+18}
{Kawash}, A.~M., {et~al.} 2018, \apj, 857, 131

\bibitem[{{Keane} {et~al.}(2018){Keane}, {Barr}, {Jameson}, {Morello}, {Caleb},
  {Bhandari}, {Petroff}, {Possenti}, {Burgay}, {Tiburzi}, {Bailes}, {Bhat},
  {Burke-Spolaor}, {Eatough}, {Flynn}, {Jankowski}, {Johnston}, {Kramer},
  {Levin}, {Ng}, {van Straten}, \& {Krishnan}}]{kbj+18}
{Keane}, E.~F., {et~al.} 2018, \mnras, 473, 116

\bibitem[{{Keith} {et~al.}(2010){Keith}, {Jameson}, {van Straten}, {Bailes},
  {Johnston}, {Kramer}, {Possenti}, {Bates}, {Bhat}, {Burgay}, {Burke-Spolaor},
  {D'Amico}, {Levin}, {McMahon}, {Milia}, \& {Stappers}}]{kjs+10}
{Keith}, M.~J., {et~al.} 2010, \mnras, 409, 619

\bibitem[{{Kondratiev} {et~al.}(2016){Kondratiev}, {Verbiest}, {Hessels},
  {Bilous}, {Stappers}, {Kramer}, {Keane}, {Noutsos}, {Os{\l}owski}, {Breton},
  {Hassall}, {Alexov}, {Cooper}, {Falcke}, {Grie{\ss}meier}, {Karastergiou},
  {Kuniyoshi}, {Pilia}, {Sobey}, {ter Veen}, {van Leeuwen}, {Weltevrede},
  {Bell}, {Broderick}, {Corbel}, {Eisl{\"o}ffel}, {Markoff}, {Rowlinson},
  {Swinbank}, {Wijers}, {Wijnands}, \& {Zarka}}]{kvh+15}
{Kondratiev}, V.~I., {et~al.} 2016, \aap, 585, A128

\bibitem[{{Kramer} {et~al.}(1999){Kramer}, {Lange}, {Lorimer}, {Backer},
  {Xilouris}, {Jessner}, \& {Wielebinski}}]{kll+99}
{Kramer}, M., {Lange}, C., {Lorimer}, D.~R., {Backer}, D.~C., {Xilouris},
  K.~M., {Jessner}, A., \& {Wielebinski}, R. 1999, \apj, 526, 957

\bibitem[{{Kramer} {et~al.}(1998){Kramer}, {Xilouris}, {Lorimer}, {Doroshenko},
  {Jessner}, {Wielebinski}, {Wolszczan}, \& {Camilo}}]{kxl+98}
{Kramer}, M., {Xilouris}, K.~M., {Lorimer}, D.~R., {Doroshenko}, O., {Jessner},
  A., {Wielebinski}, R., {Wolszczan}, A., \& {Camilo}, F. 1998, \apj, 501, 270

\bibitem[{{Kuzmin} \& {Losovsky}(2001)}]{kl01}
{Kuzmin}, A.~D., \& {Losovsky}, B.~Y. 2001, \aap, 368, 230

\bibitem[{{Lazarus} {et~al.}(2015){Lazarus}, {Brazier}, {Hessels},
  {Karako-Argaman}, {Kaspi}, {Lynch}, {Madsen}, {Patel}, {Ransom}, {Scholz},
  {Swiggum}, {Zhu}, {Allen}, {Bogdanov}, {Camilo}, {Cardoso}, {Chatterjee},
  {Cordes}, {Crawford}, {Deneva}, {Ferdman}, {Freire}, {Jenet}, {Knispel},
  {Lee}, {van Leeuwen}, {Lorimer}, {Lyne}, {McLaughlin}, {Siemens}, {Spitler},
  {Stairs}, {Stovall}, \& {Venkataraman}}]{lbh+15}
{Lazarus}, P., {et~al.} 2015, \apj, 812, 81

\bibitem[{{Levin} {et~al.}(2016){Levin}, {McLaughlin}, {Jones}, {Cordes},
  {Stinebring}, {Chatterjee}, {Dolch}, {Lam}, {Lazio}, {Palliyaguru},
  {Arzoumanian}, {Crowter}, {Demorest}, {Ellis}, {Ferdman}, {Fonseca},
  {Gonzalez}, {Jones}, {Nice}, {Pennucci}, {Ransom}, {Stairs}, {Stovall},
  {Swiggum}, \& {Zhu}}]{lmj+16}
{Levin}, L., {et~al.} 2016, \apj, 818, 166

\bibitem[{{Lewandowski} {et~al.}(2004){Lewandowski}, {Wolszczan}, {Feiler},
  {Konacki}, \& {So{\l}tysi{\'n}ski}}]{lwf+04}
{Lewandowski}, W., {Wolszczan}, A., {Feiler}, G., {Konacki}, M., \&
  {So{\l}tysi{\'n}ski}, T. 2004, \apj, 600, 905

\bibitem[{{Lommen} {et~al.}(2000){Lommen}, {Zepka}, {Backer}, {McLaughlin},
  {Cordes}, {Arzoumanian}, \& {Xilouris}}]{lzb+00}
{Lommen}, A.~N., {Zepka}, A., {Backer}, D.~C., {McLaughlin}, M., {Cordes},
  J.~M., {Arzoumanian}, Z., \& {Xilouris}, K. 2000, \apj, 545, 1007

\bibitem[{{Lorimer}(1994)}]{lor94}
{Lorimer}, D. 1994, PhD thesis, The University of Manchester

\bibitem[{{Lorimer} \& {Kramer}(2004)}]{lk04}
{Lorimer}, D.~R., \& {Kramer}, M. 2004, {Handbook of Pulsar Astronomy}

\bibitem[{{Lorimer} {et~al.}(1998){Lorimer}, {Lyne}, \& {Camilo}}]{llc98}
{Lorimer}, D.~R., {Lyne}, A.~G., \& {Camilo}, F. 1998, \aap, 331, 1002

\bibitem[{{Lorimer} {et~al.}(1995{\natexlab{a}}){Lorimer}, {Yates}, {Lyne}, \&
  {Gould}}]{lylg95}
{Lorimer}, D.~R., {Yates}, J.~A., {Lyne}, A.~G., \& {Gould}, D.~M.
  1995{\natexlab{a}}, \mnras, 273, 411

\bibitem[{{Lorimer} {et~al.}(1995{\natexlab{b}}){Lorimer}, {Nicastro}, {Lyne},
  {Bailes}, {Manchester}, {Johnston}, {Bell}, {D'Amico}, \&
  {Harrison}}]{lnl+95}
{Lorimer}, D.~R., {et~al.} 1995{\natexlab{b}}, \apj, 439, 933

\bibitem[{{Lorimer} {et~al.}(2006){Lorimer}, {Faulkner}, {Lyne}, {Manchester},
  {Kramer}, {McLaughlin}, {Hobbs}, {Possenti}, {Stairs}, {Camilo}, {Burgay},
  {D'Amico}, {Corongiu}, \& {Crawford}}]{lfl+06}
---. 2006, \mnras, 372, 777

\bibitem[{{Lorimer} {et~al.}(2015){Lorimer}, {Esposito}, {Manchester},
  {Possenti}, {Lyne}, {McLaughlin}, {Kramer}, {Hobbs}, {Stairs}, \&
  {Burgay}}]{lem+15}
---. 2015, \mnras, 450, 2185

\bibitem[{{Lynch} {et~al.}(2013){Lynch}, {Boyles}, {Ransom}, {Stairs},
  {Lorimer}, {McLaughlin}, {Hessels}, {Kaspi}, {Kondratiev}, {Archibald},
  {Berndsen}, {Cardoso}, {Cherry}, {Epstein}, {Karako-Argaman}, {McPhee},
  {Pennucci}, {Roberts}, {Stovall}, \& {van Leeuwen}}]{lbr+13}
{Lynch}, R.~S., {et~al.} 2013, \apj, 763, 81

\bibitem[{{Lynch} {et~al.}(2018){Lynch}, {Swiggum}, {Kondratiev}, {Kaplan},
  {Stovall}, {Fonseca}, {Roberts}, {Levin}, {DeCesar}, {Cui}, {Cenko},
  {Gatkine}, {Archibald}, {Banaszak}, {Biwer}, {Boyles}, {Chawla}, {Dartez},
  {Day}, {Ford}, {Flanigan}, {Hessels}, {Hinojosa}, {Jenet}, {Karako-Argaman},
  {Kaspi}, {Leake}, {Lunsford}, {Martinez}, {Mata}, {McLaughlin}, {Noori},
  {Ransom}, {Rohr}, {Siemens}, {Spiewak}, {Stairs}, {van Leeuwen}, {Walker}, \&
  {Wells}}]{lsk+18}
---. 2018, \apj, 859, 93

\bibitem[{{Lyne} {et~al.}(1998){Lyne}, {Manchester}, {Lorimer}, {Bailes},
  {D'Amico}, {Tauris}, {Johnston}, {Bell}, \& {Nicastro}}]{lml+98}
{Lyne}, A.~G., {et~al.} 1998, \mnras, 295, 743

\bibitem[{{Manchester} {et~al.}(2005){Manchester}, {Hobbs}, {Teoh}, \&
  {Hobbs}}]{mhth05}
{Manchester}, R.~N., {Hobbs}, G.~B., {Teoh}, A., \& {Hobbs}, M. 2005, \aj, 129,
  1993

\bibitem[{{Manchester} {et~al.}(1996){Manchester}, {Lyne}, {D'Amico}, {Bailes},
  {Johnston}, {Lorimer}, {Harrison}, {Nicastro}, \& {Bell}}]{mld+96}
{Manchester}, R.~N., {et~al.} 1996, \mnras, 279, 1235

\bibitem[{{Manchester} {et~al.}(2013){Manchester}, {Hobbs}, {Bailes}, {Coles},
  {van Straten}, {Keith}, {Shannon}, {Bhat}, {Brown}, {Burke-Spolaor},
  {Champion}, {Chaudhary}, {Edwards}, {Hampson}, {Hotan}, {Jameson}, {Jenet},
  {Kesteven}, {Khoo}, {Kocz}, {Maciesiak}, {Oslowski}, {Ravi}, {Reynolds},
  {Sarkissian}, {Verbiest}, {Wen}, {Wilson}, {Yardley}, {Yan}, \&
  {You}}]{mhb+13}
---. 2013, \pasa, 30, e017

\bibitem[{{McLaughlin} {et~al.}(2006){McLaughlin}, {Lyne}, {Lorimer}, {Kramer},
  {Faulkner}, {Manchester}, {Cordes}, {Camilo}, {Possenti}, {Stairs}, {Hobbs},
  {D'Amico}, {Burgay}, \& {O'Brien}}]{mll+06}
{McLaughlin}, M.~A., {et~al.} 2006, \nat, 439, 817

\bibitem[{{Mohanty}(1983)}]{moh83}
{Mohanty}, D.~K. 1983, in IAU Symposium, Vol. 101, Supernova Remnants and their
  X-ray Emission, ed. J.~{Danziger} \& P.~{Gorenstein}, 503

\bibitem[{{Murphy} {et~al.}(2017){Murphy}, {Kaplan}, {Bell}, {Callingham},
  {Croft}, {Johnston}, {Dobie}, {Zic}, {Hughes}, {Lynch}, {Hancock},
  {Hurley-Walker}, {Lenc}, {Dwarakanath}, {For}, {Gaensler}, {Hindson},
  {Johnston-Hollitt}, {Kapi{\'n}ska}, {McKinley}, {Morgan}, {Offringa},
  {Procopio}, {Staveley-Smith}, {Wayth}, {Wu}, \& {Zheng}}]{mkb+17}
{Murphy}, T., {et~al.} 2017, \pasa, 34, e020

\bibitem[{{Navarro} {et~al.}(2003){Navarro}, {Anderson}, \& {Freire}}]{naf03}
{Navarro}, J., {Anderson}, S.~B., \& {Freire}, P.~C. 2003, \apj, 594, 943

\bibitem[{{Nicastro} {et~al.}(1995){Nicastro}, {Lyne}, {Lorimer}, {Harrison},
  {Bailes}, \& {Skidmore}}]{nll+95}
{Nicastro}, L., {Lyne}, A.~G., {Lorimer}, D.~R., {Harrison}, P.~A., {Bailes},
  M., \& {Skidmore}, B.~D. 1995, \mnras, 273, L68

\bibitem[{{Nice} {et~al.}(2013){Nice}, {Altiere}, {Bogdanov}, {Cordes},
  {Farrington}, {Hessels}, {Kaspi}, {Lyne}, {Popa}, {Ransom}, {Sanpa-arsa},
  {Stappers}, {Wang}, {Allen}, {Bhat}, {Brazier}, {Camilo}, {Champion},
  {Chatterjee}, {Crawford}, {Deneva}, {Desvignes}, {Freire}, {Jenet},
  {Knispel}, {Lazarus}, {Lee}, {van Leeuwen}, {Lorimer}, {Lynch}, {McLaughlin},
  {Scholz}, {Siemens}, {Stairs}, {Stovall}, {Venkataraman}, \& {Zhu}}]{nab+13}
{Nice}, D.~J., {et~al.} 2013, \apj, 772, 50

\bibitem[{Oliphant(2006)}]{numpy}
Oliphant, T.~E. 2006, A guide to NumPy, Vol.~1 (Trelgol Publishing USA)

\bibitem[{{Price-Whelan} {et~al.}(2018){Price-Whelan}, {Sip{\H{o}}cz},
  {G{\"u}nther}, {Lim}, {Crawford}, {Conseil}, {Shupe}, {Craig}, {Dencheva},
  {Ginsburg}, {VanderPlas}, {Bradley}, {P{\'e}rez-Su{\'a}rez}, {de Val-Borro},
  {Paper Contributors}, {Aldcroft}, {Cruz}, {Robitaille}, {Tollerud},
  {Coordination Committee}, {Ardelean}, {Babej}, {Bach}, {Bachetti}, {Bakanov},
  {Bamford}, {Barentsen}, {Barmby}, {Baumbach}, {Berry}, {Biscani}, {Boquien},
  {Bostroem}, {Bouma}, {Brammer}, {Bray}, {Breytenbach}, {Buddelmeijer},
  {Burke}, {Calderone}, {Cano Rodr{\'\i}guez}, {Cara}, {Cardoso}, {Cheedella},
  {Copin}, {Corrales}, {Crichton}, {D{\textquoteright}Avella}, {Deil},
  {Depagne}, {Dietrich}, {Donath}, {Droettboom}, {Earl}, {Erben}, {Fabbro},
  {Ferreira}, {Finethy}, {Fox}, {Garrison}, {Gibbons}, {Goldstein}, {Gommers},
  {Greco}, {Greenfield}, {Groener}, {Grollier}, {Hagen}, {Hirst}, {Homeier},
  {Horton}, {Hosseinzadeh}, {Hu}, {Hunkeler}, {Ivezi{\'c}}, {Jain}, {Jenness},
  {Kanarek}, {Kendrew}, {Kern}, {Kerzendorf}, {Khvalko}, {King}, {Kirkby},
  {Kulkarni}, {Kumar}, {Lee}, {Lenz}, {Littlefair}, {Ma}, {Macleod},
  {Mastropietro}, {McCully}, {Montagnac}, {Morris}, {Mueller}, {Mumford},
  {Muna}, {Murphy}, {Nelson}, {Nguyen}, {Ninan}, {N{\"o}the}, {Ogaz}, {Oh},
  {Parejko}, {Parley}, {Pascual}, {Patil}, {Patil}, {Plunkett}, {Prochaska},
  {Rastogi}, {Reddy Janga}, {Sabater}, {Sakurikar}, {Seifert}, {Sherbert},
  {Sherwood-Taylor}, {Shih}, {Sick}, {Silbiger}, {Singanamalla}, {Singer},
  {Sladen}, {Sooley}, {Sornarajah}, {Streicher}, {Teuben}, {Thomas},
  {Tremblay}, {Turner}, {Terr{\'o}n}, {van Kerkwijk}, {de la Vega}, {Watkins},
  {Weaver}, {Whitmore}, {Woillez}, {Zabalza}, \& {Contributors}}]{astropy:2018}
{Price-Whelan}, A.~M., {et~al.} 2018, \aj, 156, 123

\bibitem[{{Qiao} {et~al.}(1995){Qiao}, {Manchester}, {Lyne}, \&
  {Gould}}]{qmlg95}
{Qiao}, G., {Manchester}, R.~N., {Lyne}, A.~G., \& {Gould}, D.~M. 1995, \mnras,
  274, 572

\bibitem[{{Ransom}(2001)}]{smr+01}
{Ransom}, S.~M. 2001, PhD thesis, Harvard University

\bibitem[{{Ransom} {et~al.}(2011){Ransom}, {Ray}, {Camilo}, {Roberts},
  {{\c{C}}elik}, {Wolff}, {Cheung}, {Kerr}, {Pennucci}, {DeCesar}, {Cognard},
  {Lyne}, {Stappers}, {Freire}, {Grove}, {Abdo}, {Desvignes}, {Donato},
  {Ferrara}, {Gehrels}, {Guillemot}, {Gwon}, {Harding}, {Johnston}, {Keith},
  {Kramer}, {Michelson}, {Parent}, {Saz Parkinson}, {Romani}, {Smith},
  {Theureau}, {Thompson}, {Weltevrede}, {Wood}, \& {Ziegler}}]{rrc+11}
{Ransom}, S.~M., {et~al.} 2011, \apjl, 727, L16

\bibitem[{{Ray} {et~al.}(1996){Ray}, {Thorsett}, {Jenet}, {van Kerkwijk},
  {Kulkarni}, {Prince}, {Sand hu}, \& {Nice}}]{rtj+96}
{Ray}, P.~S., {Thorsett}, S.~E., {Jenet}, F.~A., {van Kerkwijk}, M.~H.,
  {Kulkarni}, S.~R., {Prince}, T.~A., {Sand hu}, J.~S., \& {Nice}, D.~J. 1996,
  \apj, 470, 1103

\bibitem[{{Sanidas} {et~al.}(2019){Sanidas}, {Cooper}, {Bassa}, {Hessels},
  {Kondratiev}, {Michilli}, {Stappers}, {Tan}, {van Leeuwen}, {Cerrigone},
  {Fallows}, {Iacobelli}, {Orr{\'u}}, {Pizzo}, {Shulevski}, {Toribio}, {ter
  Veen}, {Zucca}, {Bondonneau}, {Grie{\ss}meier}, {Karastergiou}, {Kramer}, \&
  {Sobey}}]{scb+19}
{Sanidas}, S., {et~al.} 2019, \aap, 626, A104

\bibitem[{{Sayer} {et~al.}(1997){Sayer}, {Nice}, \& {Taylor}}]{snt97}
{Sayer}, R.~W., {Nice}, D.~J., \& {Taylor}, J.~H. 1997, \apj, 474, 426

\bibitem[{{Stairs} {et~al.}(1999){Stairs}, {Thorsett}, \& {Camilo}}]{stc99}
{Stairs}, I.~H., {Thorsett}, S.~E., \& {Camilo}, F. 1999, \apjs, 123, 627

\bibitem[{{Stokes} {et~al.}(1986){Stokes}, {Segelstein}, {Taylor}, \&
  {Dewey}}]{sstd86}
{Stokes}, G.~H., {Segelstein}, D.~J., {Taylor}, J.~H., \& {Dewey}, R.~J. 1986,
  \apj, 311, 694

\bibitem[{{Stovall} {et~al.}(2014){Stovall}, {Lynch}, {Ransom}, {Archibald},
  {Banaszak}, {Biwer}, {Boyles}, {Dartez}, {Day}, {Ford}, {Flanigan}, {Garcia},
  {Hessels}, {Hinojosa}, {Jenet}, {Kaplan}, {Karako-Argaman}, {Kaspi},
  {Kondratiev}, {Leake}, {Lorimer}, {Lunsford}, {Martinez}, {Mata},
  {McLaughlin}, {Roberts}, {Rohr}, {Siemens}, {Stairs}, {van Leeuwen},
  {Walker}, \& {Wells}}]{slr+14}
{Stovall}, K., {et~al.} 2014, \apj, 791, 67

\bibitem[{{Stovall} {et~al.}(2015){Stovall}, {Ray}, {Blythe}, {Dowell},
  {Eftekhari}, {Garcia}, {Lazio}, {McCrackan}, {Schinzel}, \&
  {Taylor}}]{srb+15}
---. 2015, \apj, 808, 156

\bibitem[{{Surnis} {et~al.}(2019){Surnis}, {Joshi}, {McLaughlin},
  {Krishnakumar}, {Manoharan}, \& {Naidu}}]{sjm+19}
{Surnis}, M.~P., {Joshi}, B.~C., {McLaughlin}, M.~A., {Krishnakumar}, M.~A.,
  {Manoharan}, P.~K., \& {Naidu}, A. 2019, \apj, 870, 8

\bibitem[{{Swiggum} {et~al.}(2014){Swiggum}, {Lorimer}, {McLaughlin}, {Bates},
  {Champion}, {Ransom}, {Lazarus}, {Brazier}, {Hessels}, \& {Nice}}]{slm+14}
{Swiggum}, J.~K., {et~al.} 2014, \apj, 787, 137

\bibitem[{{Swiggum} {et~al.}(2017){Swiggum}, {Kaplan}, {McLaughlin}, {Lorimer},
  {Bogdanov}, {Ray}, {Lynch}, {Gentile}, {Rosen}, {Heatherly}, {Barlow},
  {Hegedus}, {Vasquez Soto}, {Clancy}, {Kondratiev}, {Stovall}, {Istrate},
  {Penprase}, \& {Bellm}}]{skm+17}
---. 2017, \apj, 847, 25

\bibitem[{{Taylor} {et~al.}(1993){Taylor}, {Manchester}, \& {Lyne}}]{tml93}
{Taylor}, J.~H., {Manchester}, R.~N., \& {Lyne}, A.~G. 1993, \apjs, 88, 529

\bibitem[{{Toscano} {et~al.}(1998){Toscano}, {Bailes}, {Manchester}, \& {Sand
  hu}}]{tbms98}
{Toscano}, M., {Bailes}, M., {Manchester}, R.~N., \& {Sand hu}, J.~S. 1998,
  \apj, 506, 863

\bibitem[{{Tyul'bashev} {et~al.}(2016){Tyul'bashev}, {Tyul'bashev}, {Oreshko},
  \& {Logvinenko}}]{ttol16}
{Tyul'bashev}, S.~A., {Tyul'bashev}, V.~S., {Oreshko}, V.~V., \& {Logvinenko},
  S.~V. 2016, Astronomy Reports, 60, 220

\bibitem[{{van Straten} \& {Bailes}(2011)}]{sb11}
{van Straten}, W., \& {Bailes}, M. 2011, \pasa, 28, 1

\bibitem[{{Wolszczan} \& {Frail}(1992)}]{wf92}
{Wolszczan}, A., \& {Frail}, D.~A. 1992, \nat, 355, 145

\bibitem[{{Xue} {et~al.}(2017){Xue}, {Bhat}, {Tremblay}, {Ord}, {Sobey},
  {Swainston}, {Kaplan}, {Johnston}, {Meyers}, \& {McSweeney}}]{xbt+17}
{Xue}, M., {et~al.} 2017, \pasa, 34, e070

\end{thebibliography}

\begin{figure*}
    \centering
    \includegraphics{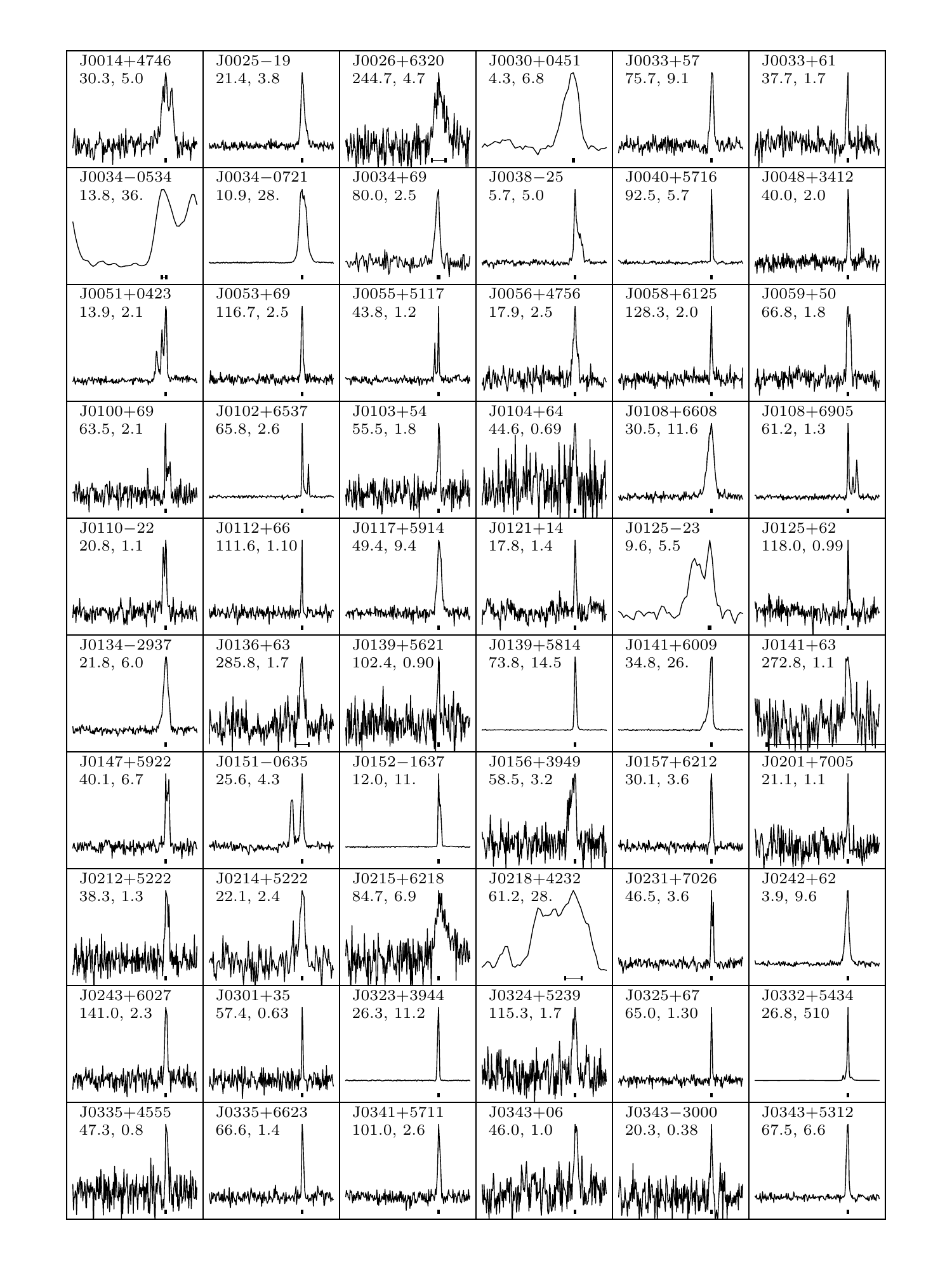}
    \caption{Pulse profiles for all detections. Text in each plot gives the pulsar name, dispersion measure in pc\,cm$^{-3}$, and flux density in mJy. Centered beneath the profiles' peaks are error bars corresponding to the expected dispersive smearing of the pulse.}
    \label{fig:prof0}
\end{figure*}

\begin{figure*}
    \centering
    \includegraphics{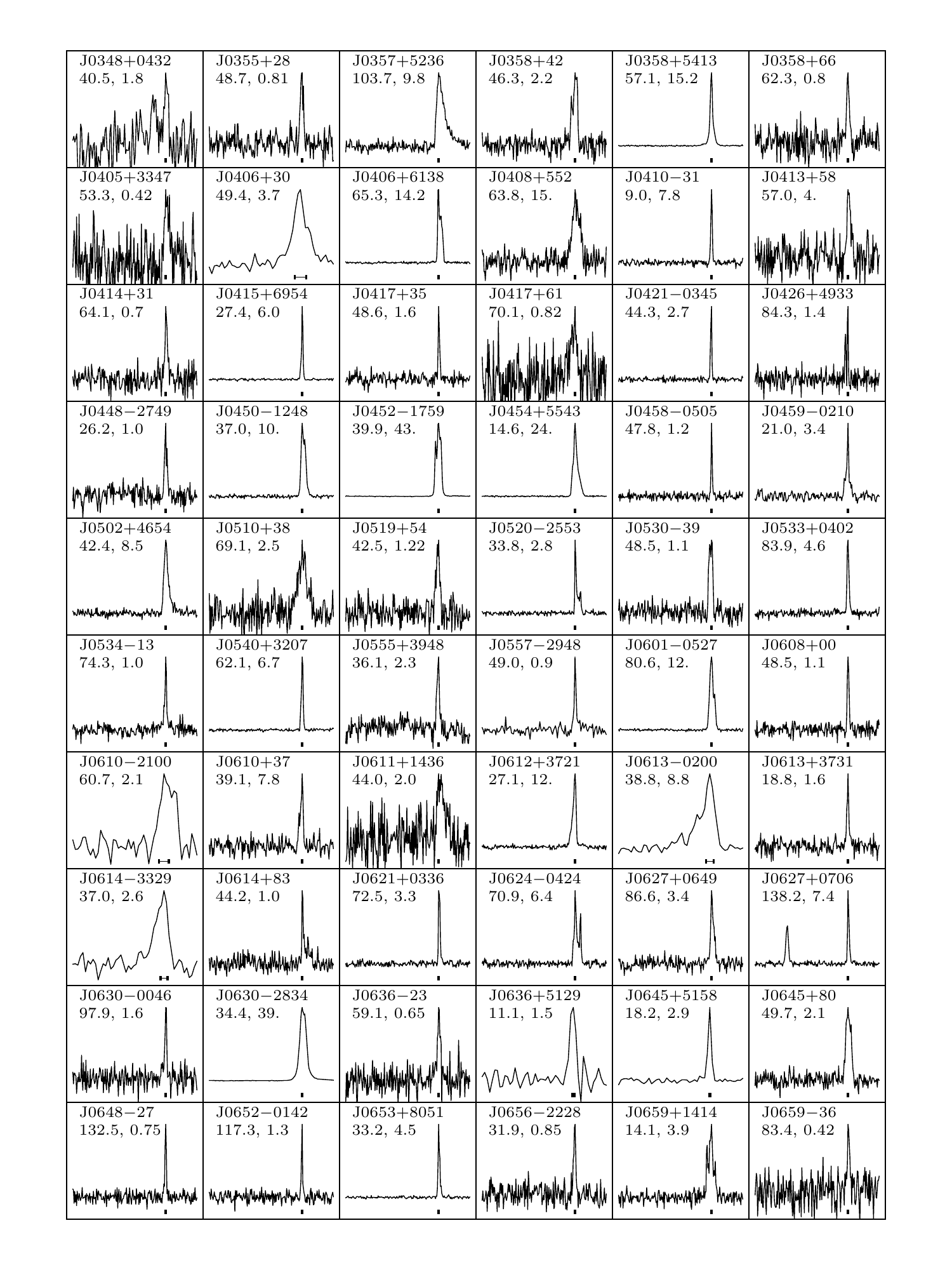}
    \caption{Profile plots (continued). See Figure \ref{fig:prof0} for details.}
    \label{fig:prof1}
\end{figure*}

\begin{figure*}
    \centering
    \includegraphics{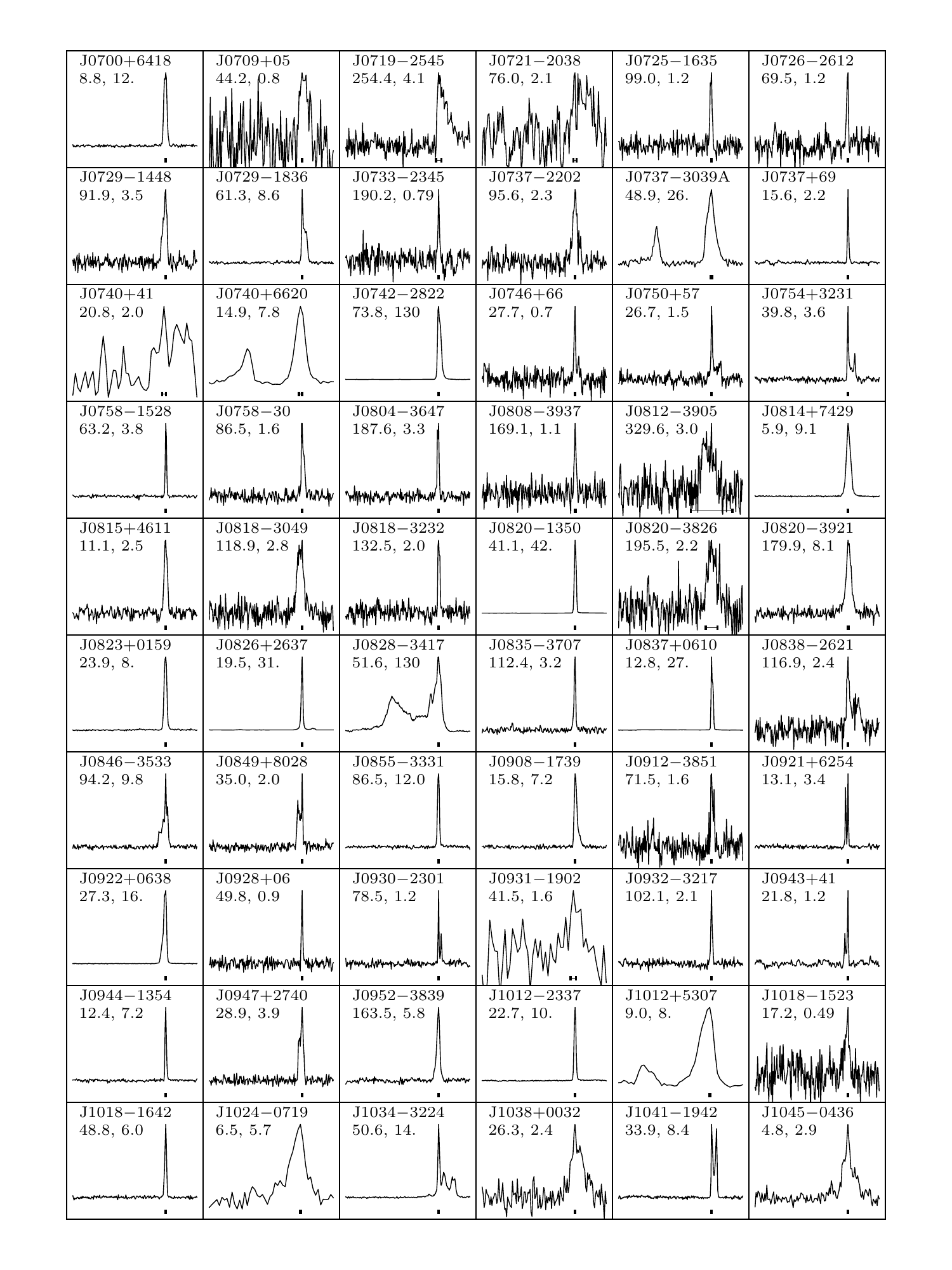}
    \caption{Profile plots (continued). See Figure \ref{fig:prof0} for details.}
    \label{fig:prof2}
\end{figure*}

\begin{figure*}
    \centering
    \includegraphics{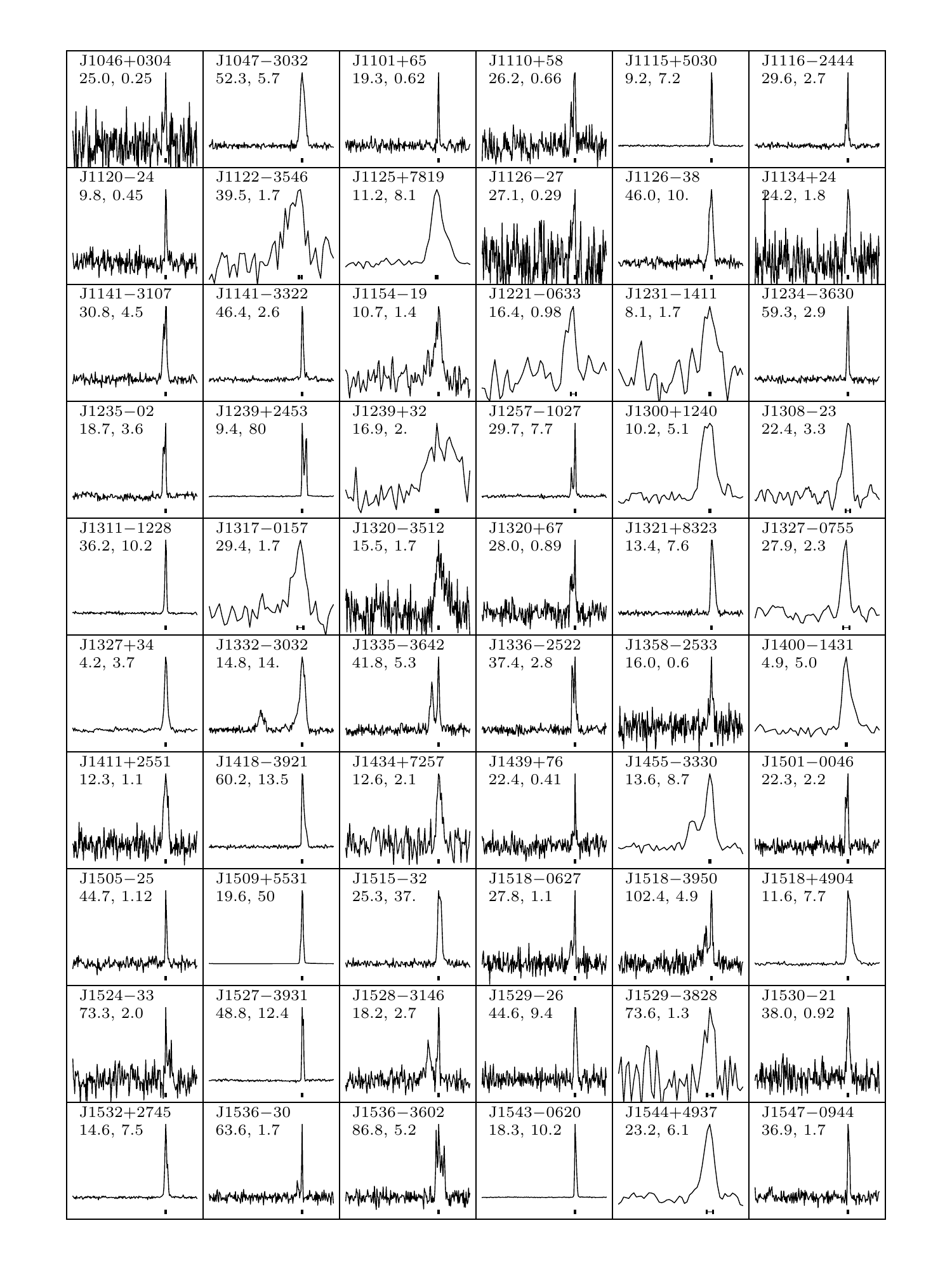}
    \caption{Profile plots (continued). See Figure \ref{fig:prof0} for details.}
    \label{fig:prof3}
\end{figure*}

\begin{figure*}
    \centering
    \includegraphics{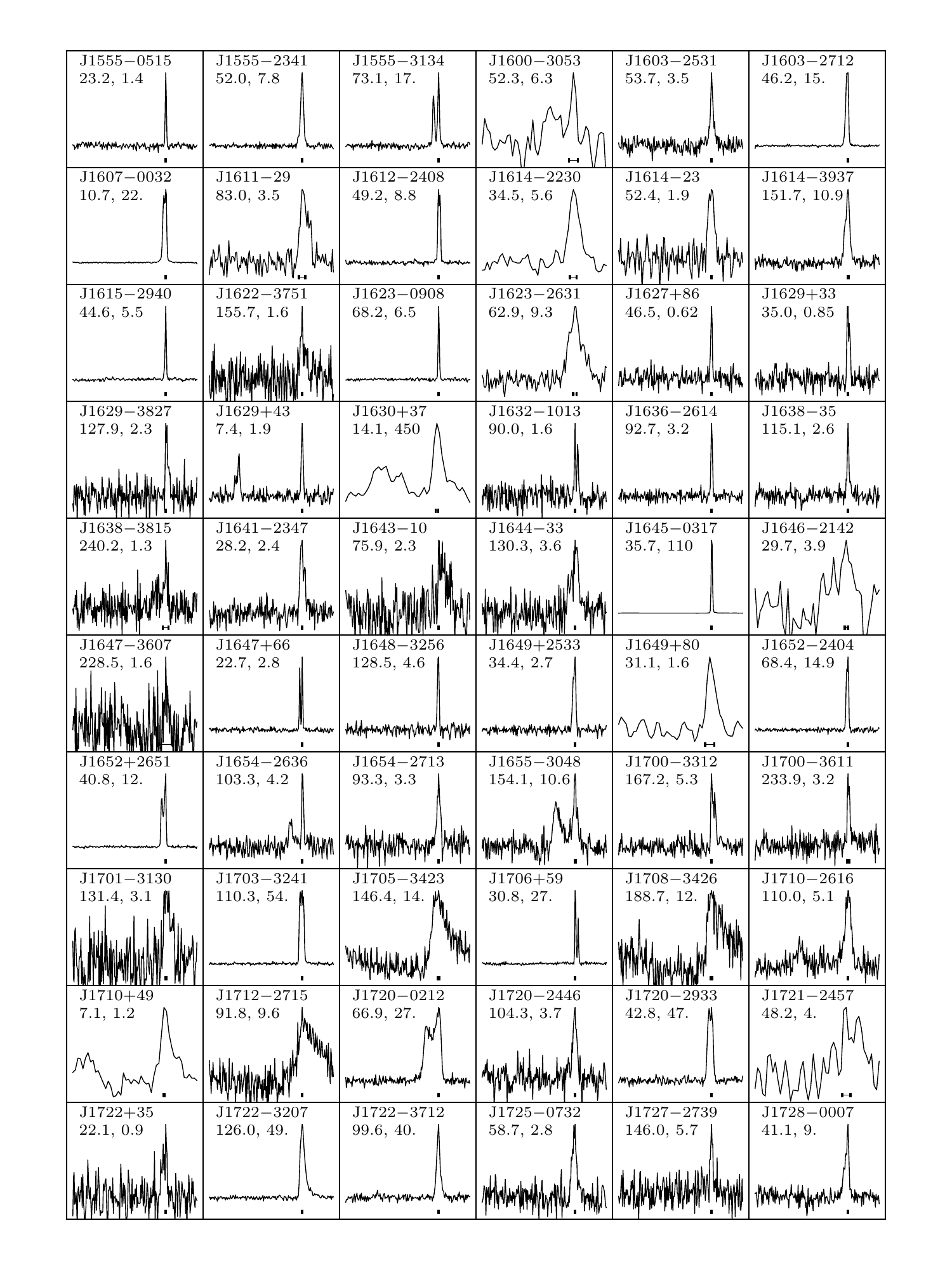}
    \caption{Profile plots (continued). See Figure \ref{fig:prof0} for details.}
    \label{fig:prof4}
\end{figure*}

\begin{figure*}
    \centering
    \includegraphics{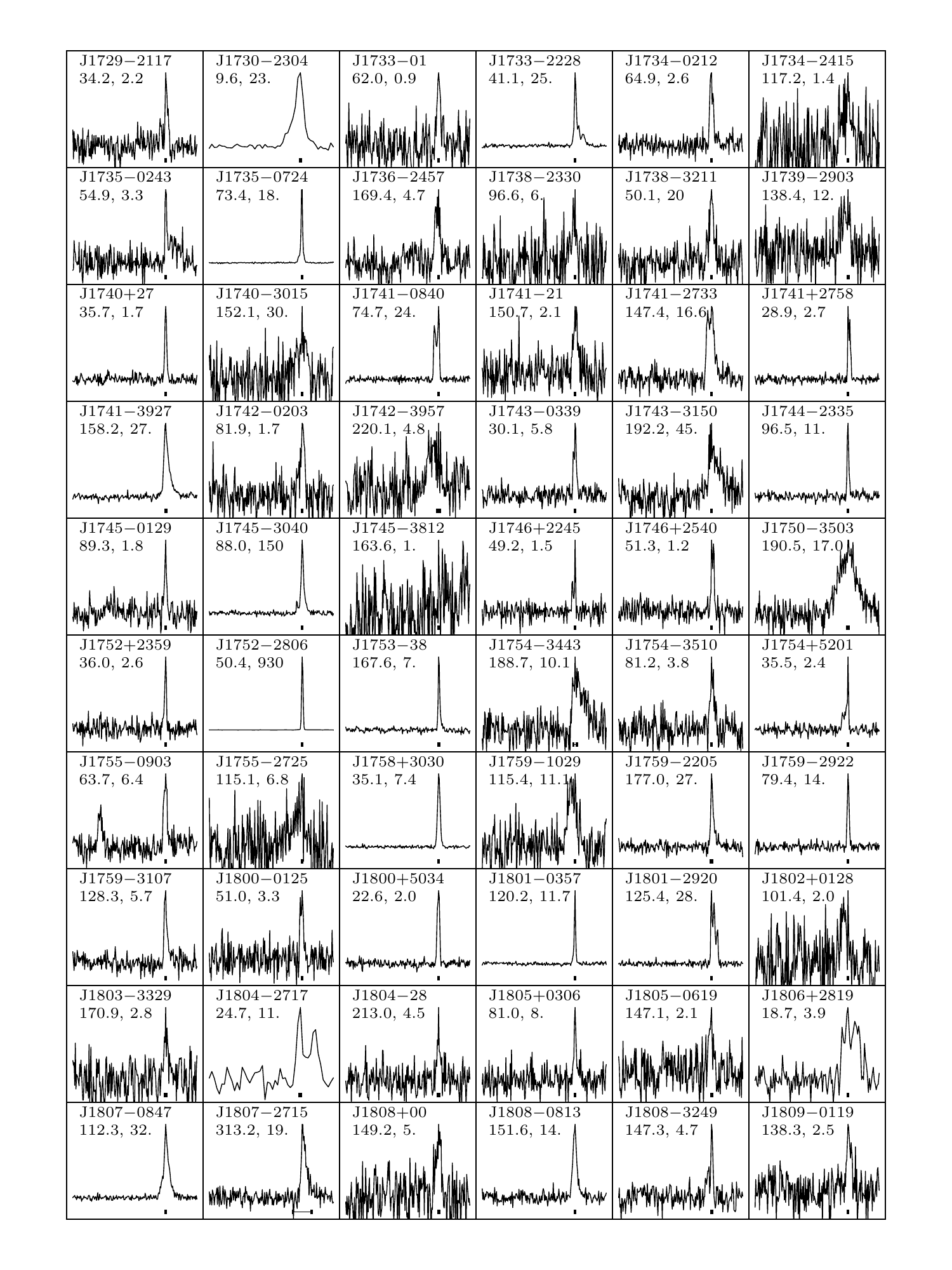}
    \caption{Profile plots (continued). See Figure \ref{fig:prof0} for details.}
    \label{fig:prof5}
\end{figure*}

\begin{figure*}
    \centering
    \includegraphics{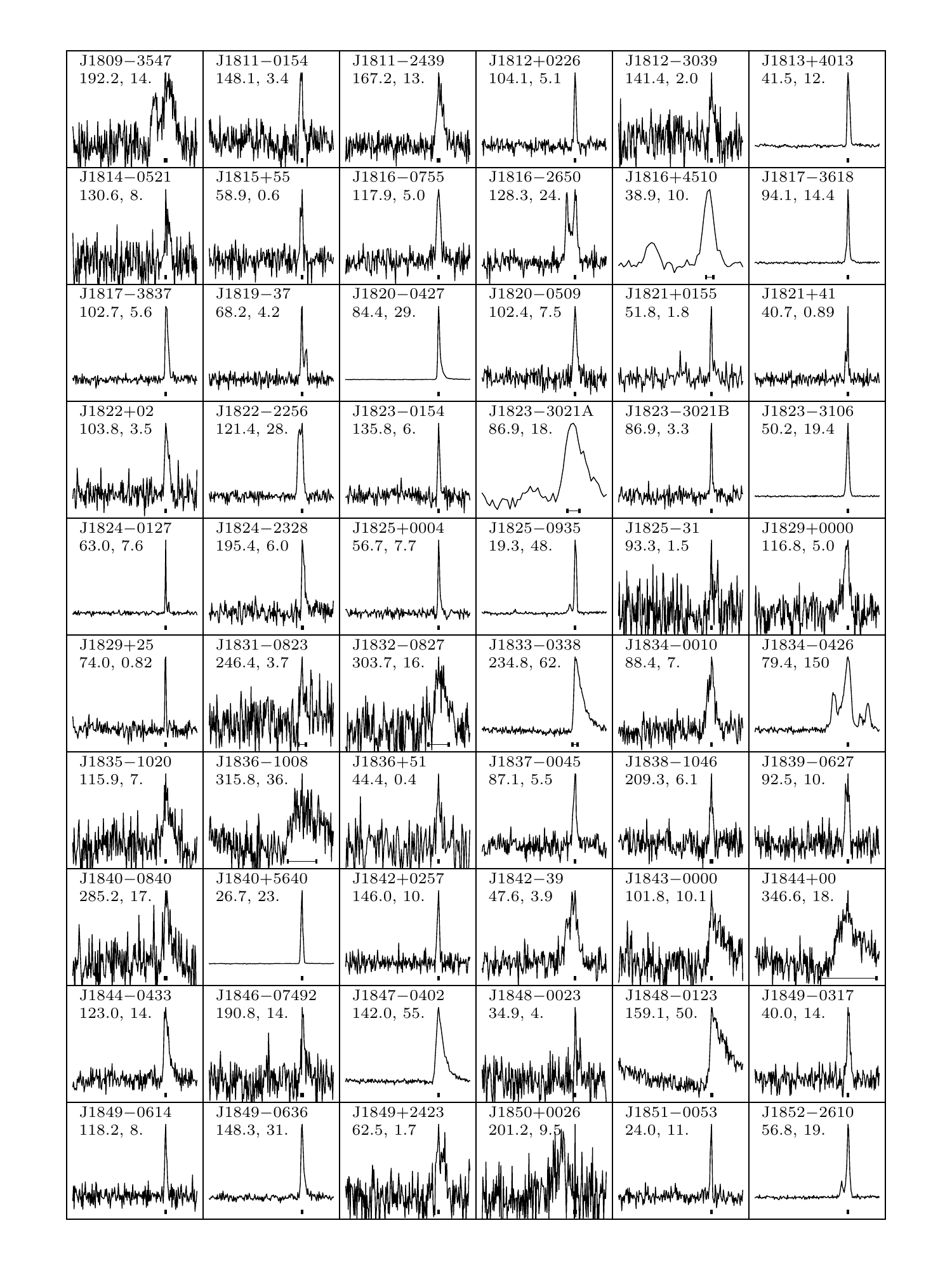}
    \caption{Profile plots (continued). See Figure \ref{fig:prof0} for details.}
    \label{fig:prof6}
\end{figure*}

\begin{figure*}
    \centering
    \includegraphics{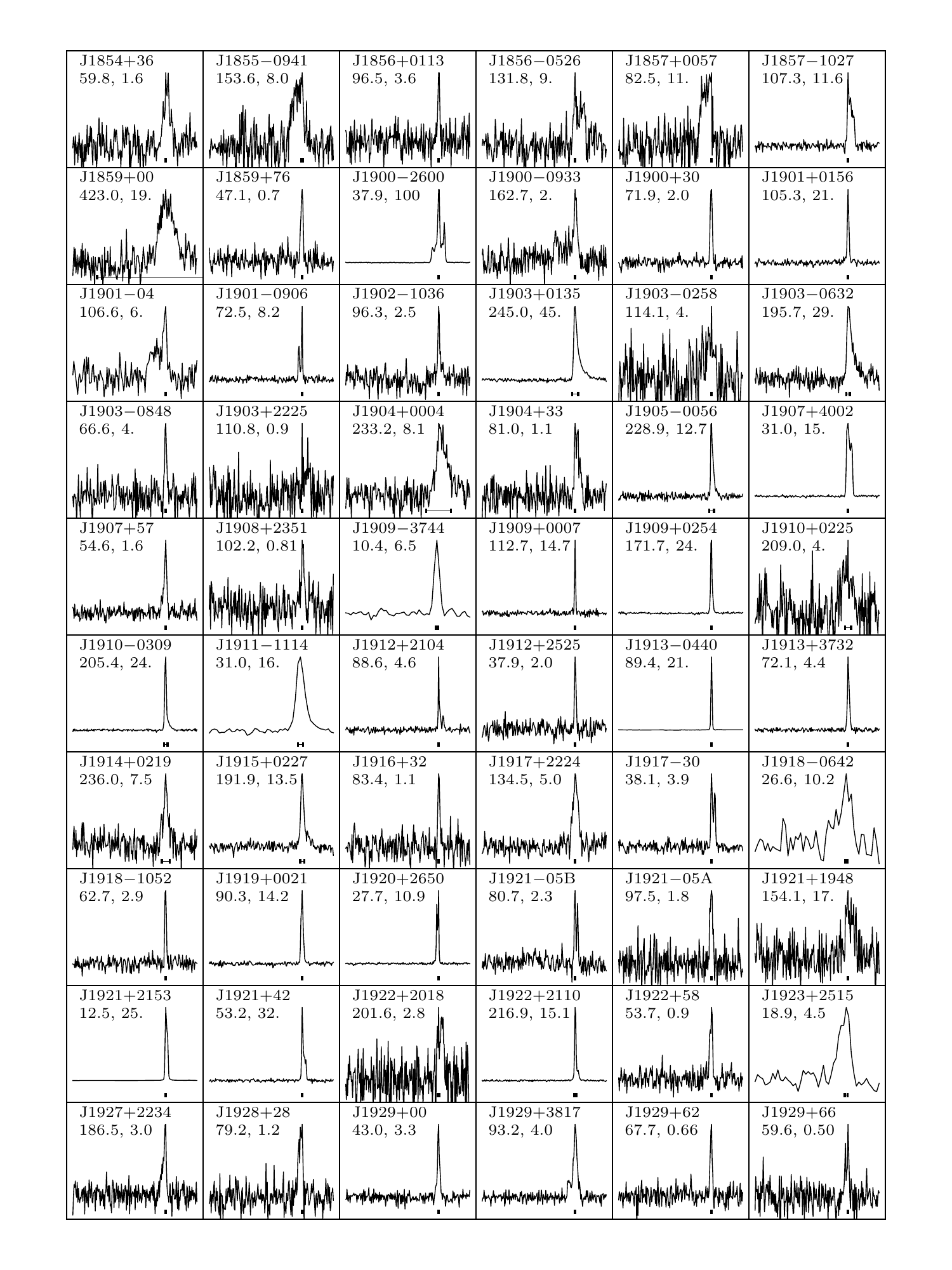}
    \caption{Profile plots (continued). See Figure \ref{fig:prof0} for details.}
    \label{fig:prof7}
\end{figure*}

\begin{figure*}
    \centering
    \includegraphics{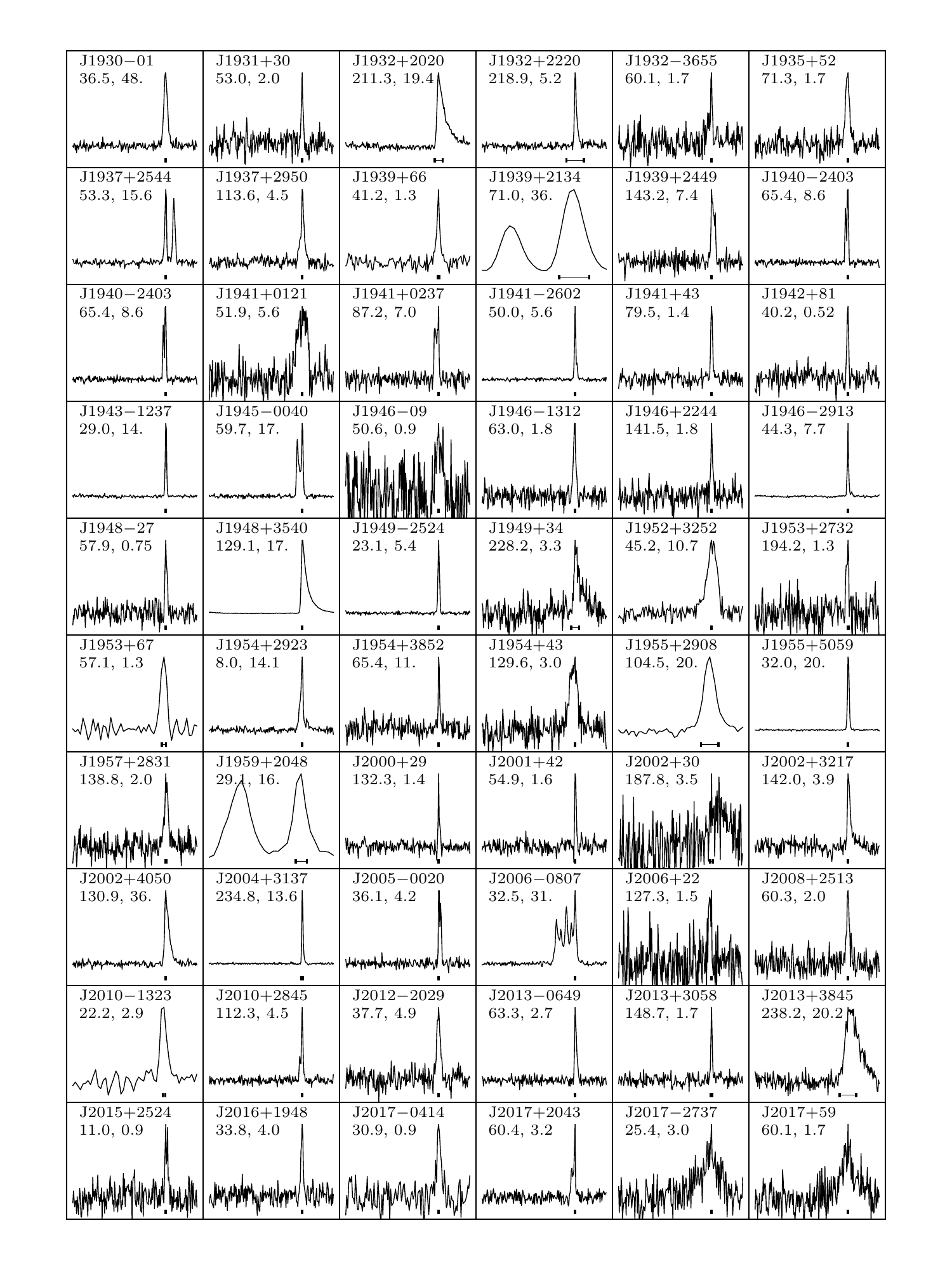}
    \caption{Profile plots (continued). See Figure \ref{fig:prof0} for details.}
    \label{fig:prof8}
\end{figure*}

\begin{figure*}
    \centering
    \includegraphics{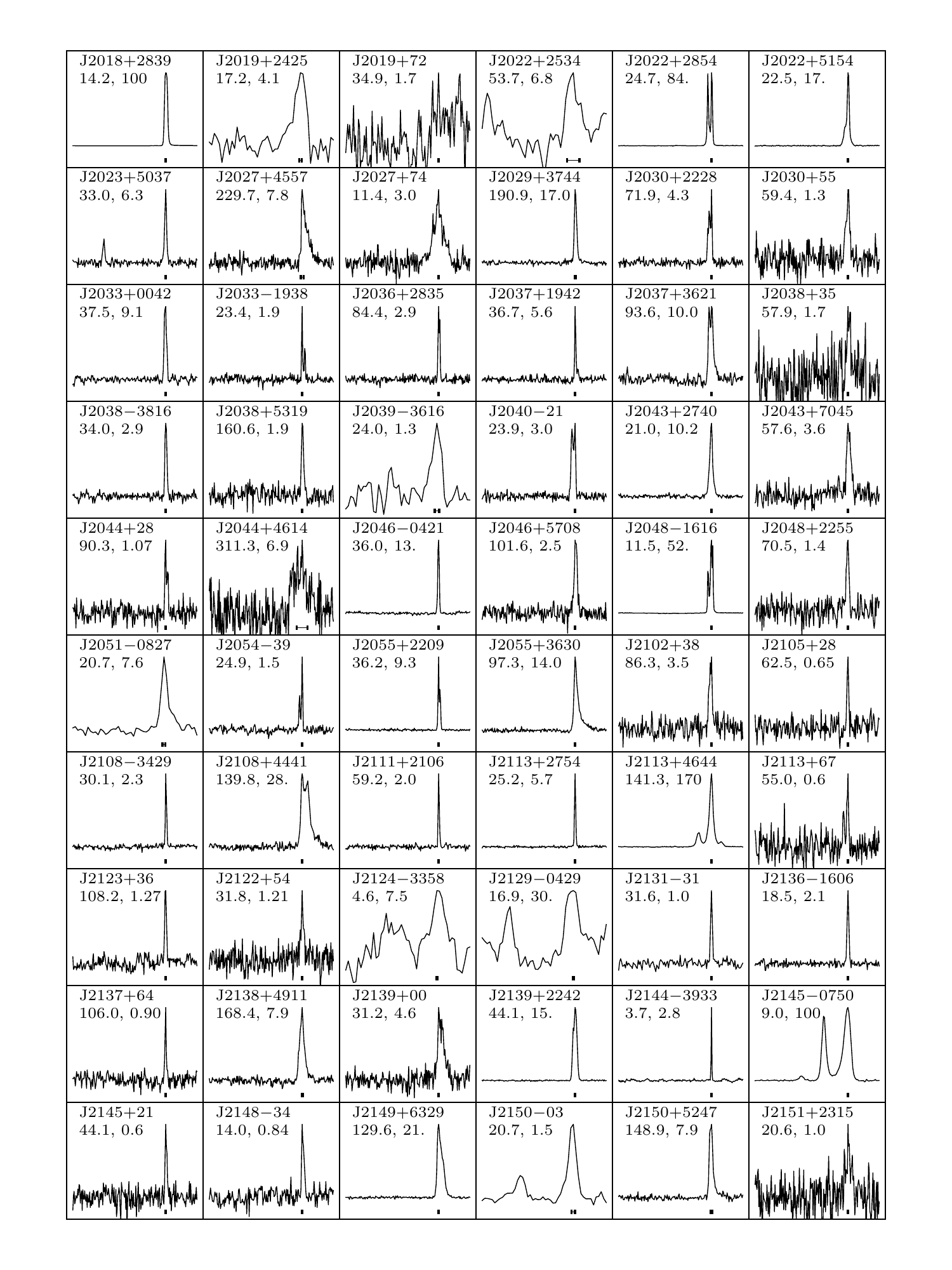}
    \caption{Profile plots (continued). See Figure \ref{fig:prof0} for details.}
    \label{fig:prof9}
\end{figure*}

\begin{figure*}
    \centering
    \includegraphics{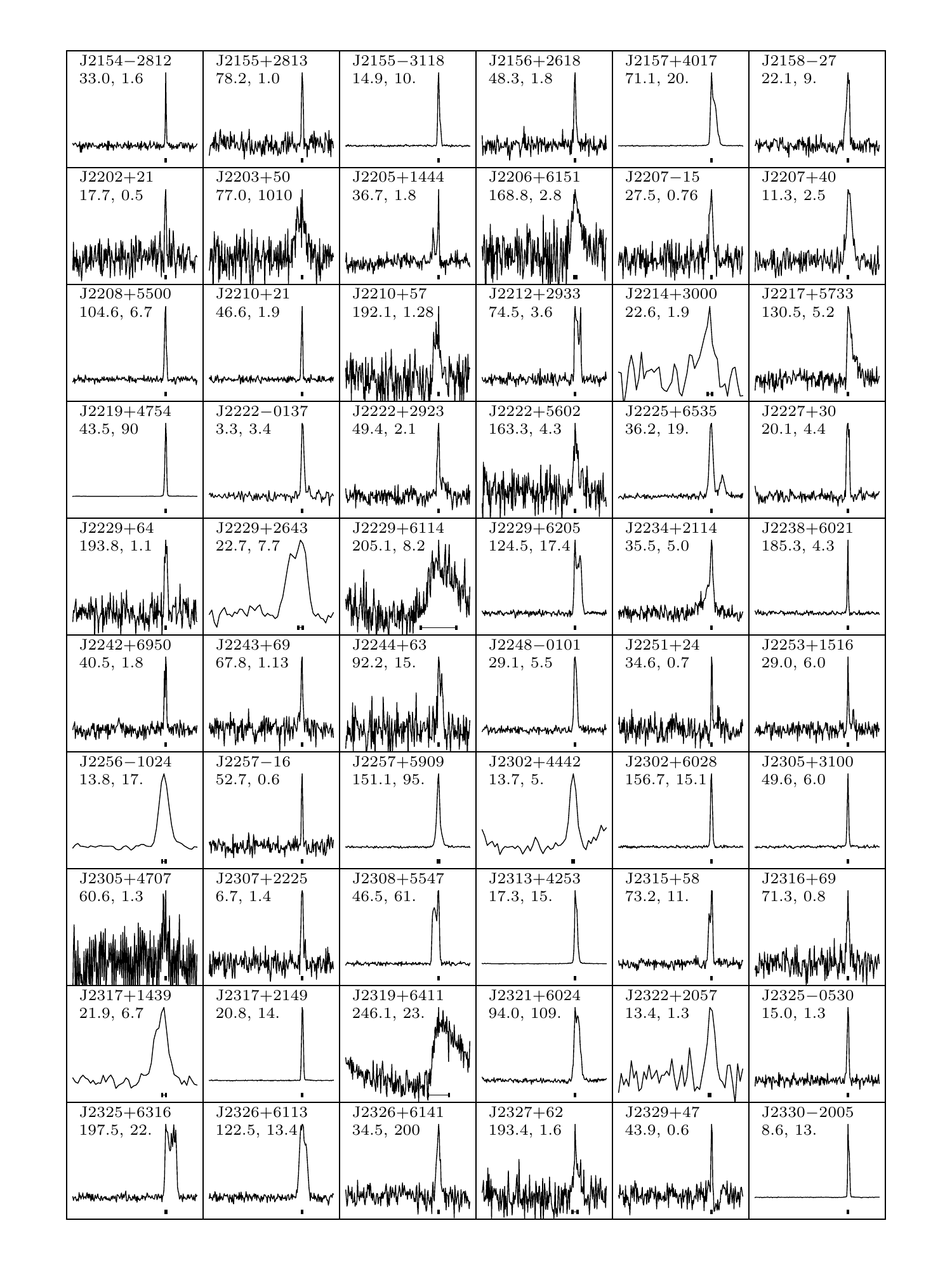}
    \caption{Profile plots (continued). See Figure \ref{fig:prof0} for details.}
    \label{fig:prof10}
\end{figure*}

\begin{figure*}
    \centering
    \includegraphics{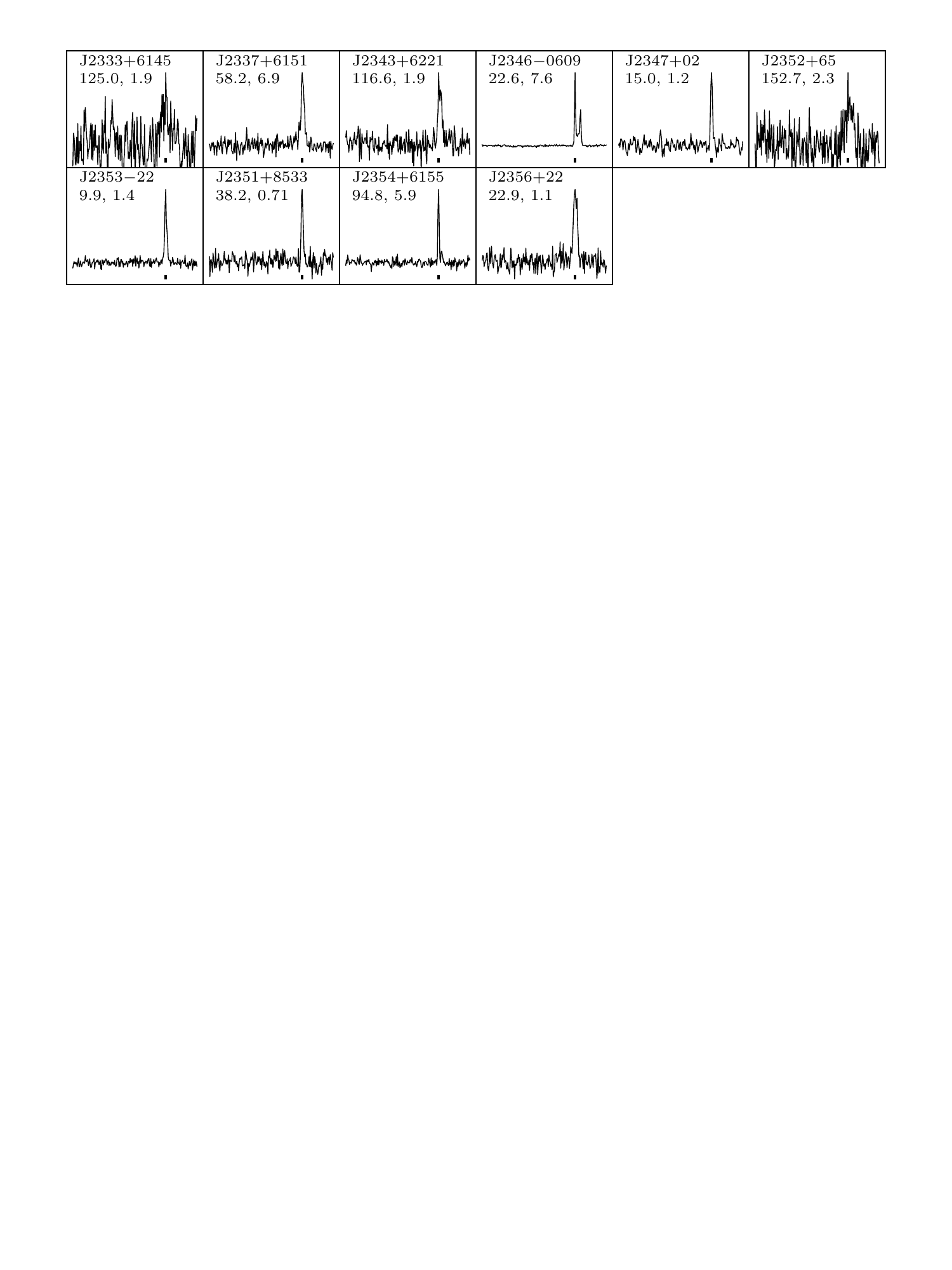}
    \caption{Profile plots (continued). See Figure \ref{fig:prof0} for details.}
    \label{fig:prof11}
\end{figure*}

\appendix
\label{app}
In Table \ref{tbl:det} we list the measured quantities of DM, pulse width, S/N, $S_{350}$, and $\alpha$. We include the references to papers from which measurements of flux density at other frequencies were taken to determine $\alpha$ in the table footnotes. Pulse profiles are shown in Figures~\ref{fig:prof0}--\ref{fig:prof11}. 
\startlongtable
\begin{deluxetable*}{lllcclllllc}
\tablewidth{0pt}
\tablecaption{Pulsar Detections in the GBNCC Survey.\label{tbl:det}}
\tablehead{
\colhead{PSR\tablenotemark{a}} & \colhead{PSR B} & \colhead{DM} & \colhead{MJD} & \colhead{$\theta$} & \colhead{$W_{50}$} & \colhead{$W_{10}$} & \colhead{S/N} & \colhead{$S_{350}$} & \colhead{$\alpha$} & \colhead{Refs.}\\
& & \colhead{(pc/cm$^3$)} &  & \colhead{(deg)} &  \colhead{(ms)} & \colhead{(ms)} &  & \colhead{(mJy)} &
}
\startdata
\object[PSR J0014+4746]{J0014+4746*} & B0011+47 & \phn30.2(3) & 55538 & 0.227 & 109 & 175 & \phn\phn29.6(5) & \phn\phn\phn5.0(8) & $-$1.0(7) & 1,2 \\ 
\object[PSR J0025-19]{J0025$-$19$^{\rm{B}}$} & \nodata & \phn21.3(3) & 58079 & 0.002 & \phn 39 & \phn 95 & \phn\phn51.3(15) & \phn\phn\phn3.8(7) & \nodata & \nodata \\ 
\object[PSR J0026+6320]{J0026+6320} & \nodata & 244.7(8) & 55198 & 0.196 & \phn 23 & \phn 48 & \phn\phn16.6(9) & \phn\phn\phn4.7(5) & $-$1.12(9) & 3,4 \\ 
\object[PSR J0030+0451]{J0030+0451} & \nodata & \phn\phn4.335(5) & 58229 & 0.086 & \phn \phn 0.7 & \phn \phn 1 & \phn\phn41.2(13) & \phn\phn\phn6.8(14) & $-$2.4(5) & 5,6,7 \\ 
\object[PSR J0033+57]{J0033+57} & \nodata & \phn75.65(8) & 55325 & 0.389 & \phn \phn 8 & \phn 18 & \phn\phn29.4(15) & \phn\phn\phn9.1(12) & \nodata & \nodata \\ 
\object[PSR J0033+61]{J0033+61} & \nodata & \phn37.6(2) & 55249 & 0.303 & \phn 16 & \phn 25 & \phn\phn12.(2) & \phn\phn\phn1.7(4) & \nodata & \nodata \\ 
\object[PSR J0034-0534]{J0034$-$0534} & \nodata & \phn13.76(19) & 57380 & 0.253 & \phn \phn 0.7 & \phn \phn 0.8 & \phn103.7(18) & \phn\phn36.(7) & \phn\phn\dag & \nodata \\ 
\object[PSR J0034-0721]{J0034$-$0721*} & B0031$-$07 & \phn10.9(2) & 57387 & 0.277 & \phn 62 & 106 & \phn184.(8) & \phn\phn28.(5) & $-$2.4(4) & 8,9,10,2,11,12 \\
\enddata
\end{deluxetable*} 
Here we list pulsars for which we have measured DM to have changed from previous measurements by $\geq$3$\sigma$. 
\startlongtable
\begin{deluxetable*}{lllc}
\tablecaption{Pulsars with $\geq 3\sigma$ DM changes.\label{tbl:dm}}
\tablehead{\colhead{Name} & \colhead{DM$_{\rm cat}$\tablenotemark{a}} & \colhead{DM$_{\rm search}$\tablenotemark{b}} & \colhead{Period}  \\
& \colhead{(pc\,cm$^{-3}$)} & \colhead{(pc\,cm$^{-3}$)} & \colhead{(s)} }
\startdata
J0026+6320 & 245.06(6) & 244.70(8) & 0.318 \\ 
J0218+4232 & \phn61.252(5) & \phn61.230(2) & 0.002 \\ 
J0502+4654 & \phn41.83(2) & \phn42.38(16) & 0.639 \\ 
J0610$-$2100 & \phn60.6662(17) & \phn60.700(3) & 0.004 \\ 
J0610+37 & \phn27.1549(3) & \phn39.09(11) & 0.444 \\ 
J0740+6620 & \phn14.9617(2) & \phn14.950(2) & 0.003 \\ 
J0818$-$3049 & 133.7(2) & 118.9(2) & 0.764 \\ 
J1125+7819 & \phn11.73(15) & \phn11.220(4) & 0.004 \\
\enddata
\tablenotetext{a}{DM value from ATNF catalog.}
\tablenotetext{b}{DM value from our search.}
\end{deluxetable*}

\end{document}